\documentclass[11pt,a4paper]{article}
\usepackage[a4paper,margin=1in]{geometry}
\usepackage{jheppub}
\usepackage[T1]{fontenc}
\usepackage[english]{babel}
\usepackage[usenames,dvipsnames]{xcolor}
\usepackage{comment}
\usepackage{float,graphicx,subfigure,multicol,multirow}
\usepackage{indentfirst,enumitem,multirow,appendix,caption,amsmath,amssymb,mathtools,braket,bm,bbm,nicefrac,stackrel}

\let\AA\undefined

\newcommand{\re}{\, \text{Re} \,}
\newcommand{\mt}{m_{\text{t}}}
\newcommand{\mz}{m_{\text{Z}}}
\newcommand{\cW}{c_{\text{\tiny W}}}
\newcommand{\sW}{s_{\text{\tiny W}}}
\newcommand{\Qt}{Q_{\text{t}}}
\newcommand{\gVt}{g_{\text{Vt}}}
\newcommand{\gAt}{g_{\text{At}}}
\newcommand{\Ql}{Q_{\ell}}
\newcommand{\gVl}{g_{\text{V}\ell}}
\newcommand{\gAl}{g_{\text{A}\ell}}
\newcommand{\DZ}{D_Z}
\newcommand{\VV}{\text{VV}}
\newcommand{\VA}{\text{VA}}
\newcommand{\AV}{\text{AV}}
\newcommand{\AA}{\text{AA}}
\newcommand{\DD}{\text{DD}}
\newcommand{\AD}{\text{AD}}
\newcommand{\VD}{\text{VD}}
\newcommand{\pV}{{\phi \text{V}}}
\newcommand{\pA}{{\phi \text{A}}}
\newcommand{\tA}{{\text{t} \gamma}}
\newcommand{\tZ}{{\text{t} Z}}

\title{Quantum tops at circular lepton colliders}

\author[a,b,c]{Fabio Maltoni,}
\author[d]{Claudio Severi,}
\author[c]{Simone Tentori,}
\author[d]{Eleni Vryonidou}

\affiliation[a]{Dipartimento di Fisica e Astronomia, Universit\`a di Bologna, via Irnerio 46, 40126 Bologna, Italy}
\affiliation[b]{INFN, Sezione di Bologna, Bologna, Italy}
\affiliation[c]{Center for Cosmology, Particle Physics and Phenomenology, Université Catholique de Louvain, Louvain-la-Neuve, Belgium}
\affiliation[d]{Department of Physics and Astronomy, University of Manchester, Oxford Road, Manchester M13~9PL, United Kingdom}

\emailAdd{fabio.maltoni@unibo.it}
\emailAdd{claudio.severi@manchester.ac.uk}
\emailAdd{simone.tentori@uclouvain.be}
\emailAdd{eleni.vryonidou@manchester.ac.uk}

\preprint{
\begin{flushright}
\end{flushright}
}

\abstract{We study the quantum properties of top quark pairs in lepton colliders with unpolarised beams, including spin correlations, entanglement, and violation of Bell inequalities. We present analytical results in the SM and in the SMEFT and discuss several practical aspects, like the choice of quantisation axes and $t \bar t$ threshold effects. We also note a correspondence between parity symmetry and entanglement. We find that quantum observables exhibit a rich phenomenology in the SM, and can also provide additional leverage in detecting new physics residing at higher scales. }
\keywords{}

\begin{document}

\maketitle

\clearpage

\section{Introduction}
The top quark is the only elementary particle known with a mass almost exactly equal to one unit of $v/\sqrt{2}$, where $v= 246$ GeV represents the vacuum expectation value of the Higgs field. In this regard, it is the only fermion acquiring a natural mass through the Standard Model (SM) mechanism of electroweak symmetry breaking. This straightforward observation has sparked numerous theoretical speculations regarding the nature and role of the top quark in fundamental interactions, both within the SM and beyond, ever since the first evidence for its existence gathered 30 years ago at FNAL~\cite{CDF:1994juo}. 

Thanks to initial measurements in $p\bar p$ collisions and the extensive $pp$ data collected at the LHC, we have accumulated increasingly precise information on the unique properties of the top quark~\cite{ParticleDataGroup:2022pth}. Its mass is now known with precision at the per mille level \cite{CMS:2023wnd}. Its strong and electroweak charges/couplings have also been determined to a high degree of accuracy. Furthermore, its average lifetime has been confirmed to be slightly shorter than the timescale required for strong interactions to bind it into a meson, and certainly much shorter than the timescale needed to affect its spin. So it happens that at a hadron collider with sufficient energy, top quarks are produced in pairs via strong interactions, but subsequently decay weakly, transferring all their quantum numbers intact to the $bW$ final states. These decay products offer a very clear signature for studying these events.

Interestingly, at a lepton collider, the top quark electroweak (EW) properties dominate the phenomenology. In such a clean environment, characterised by low QCD radiation, we will have access not only to EW properties but also to subtle QCD effects that only come into play at higher orders. As of now, no lepton collider has reached the necessary energy to produce a pair of top and anti-top quarks. However, the situation is expected to change in the coming decades with the construction of sufficiently energetic lepton accelerators. Among the most promising proposals are circular lepton colliders that explore phenomena at the weak scale, with center-of-mass energies ranging from the $Z$ pole up to $t \bar t$ threshold, and possibly beyond. According to current designs, such machines would provide samples of approximately one million $t\bar{t}$ pairs close to the production threshold. While this number alone may not be comparable to the rate at a hadron collider (where more than three billion $t\bar{t}$ events are anticipated to be on tape by the end of the High Luminosity LHC Run), the clean environment of a lepton collider together with very precise theoretical predictions~\cite{Beneke:2015kwa} will enable measurements of a well-defined top-quark mass and width at an unprecedented level~\cite{Behnke:2013xla,CLICdp:2018esa,FCC:2018byv}, orders of magnitude better that what could ever be achievable with a proton-proton machine, regardless of the amount of integrated luminosity. Therefore, it is natural to contemplate what additional insights we could gain into the properties and couplings of the top quark at lepton colliders after the HL-LHC era~\cite{Janot:2015mqv,Durieux:2018ekg,Vryonidou:2018eyv,Durieux:2018ggn,Durieux:2018tev,Durieux:2019rbz,Jung:2020uzh,Bernardi:2022hny, deBlas:2022ofj,Durieux:2022cvf,Banelli:2020iau}.

An interesting new avenue towards the exploration of top properties is through quantum observables, an approach gaining significant momentum in the past few years. In QFT, quantum mechanics is combined with special relativity, which allows us to make (perturbative) predictions for the outcomes of particle collisions.  However, thus far, collider measurements have not predominantly focused on observables that are directly sensitive to the quantum behavior of elementary particles and their interactions. One such example is entanglement, a phenomenon central to quantum theory that lacks a classical analogue. The first observations of entanglement in particle physics have been realised in flavor measurements. The observed mixing of $K$-mesons~\cite{PhysRev.103.1901}, $D$-mesons~\cite{BaBar:2007kib,  CDF:2007bdz,LHCb:2012zll}, and $B$-mesons~\cite{ARGUS:1987xtv,D0:2006oeb,CDF:2006imy,Belle:2002bwy,LHCb:2013lrq} represents an inherently quantum mechanical behavior associated with the existence of entanglement. Spin-spin correlations in $t \bar t$ final states have also been measured at the LHC~\cite{ATLAS:2014aus, CMS:2015cal, CMS:2016piu, ATLAS:2016bac,  ATLAS:2019hau, ATLAS:2019zrq,CMS:2019nrx}. However these measurements did not distinguish between quantum and classical correlations. The ATLAS and CMS collaborations have recently achieved the detection of entanglement in $t \bar t$ pairs~\cite{ATLAS:2023fsd,CMS:2024pts,CMS:2024vqh}, paving the way for TeV scale quantum observables measurements. This endeavour has recently attracted considerable phenomenological attention~\cite{Maltoni:2024tul,Barr:2024djo,Aoude:2022imd,Severi:2022qjy,Fabbrichesi:2024xtq,Fabbrichesi:2022ovb,Altakach:2022ywa,Aoude:2023hxv,Bernal:2023ruk,Fabbrichesi:2023jep}, with studies exploring the prospects of making measurements of quantum observables and detecting entanglement in various final states, including top pair and diboson production. 

One can speculate that at very small
distances QM could be modified, and imagine scenarios where quantum properties such as entanglement would manifest themselves in ways different from those predicted by QM/QFT. This would be a fundamental change of paradigm, and fundamental tests of QM at short distance, such as those recently carried in the top sector by the ATLAS and CMS Collaborations, may be intended as early studies in this direction.  However, the focus in this work is different, and more modest: we study whether quantum observables can give us some extra leverage in sensitivity in the search of ``normal'' new physics, {\it i.e.}\@ new particles and interactions (obeying QM). The merit of carrying out quantum measurements at colliders can then be considered two-fold, one in the search of new physics within QM, and the other in the test of QM itself. Several exploratory studies have investigated the use of quantum observables to constrain new physics through LHC measurements, both in simplified models~\cite{Maltoni:2024tul} and more extensively within the SMEFT framework~\cite{Aebischer:2018csl,Charles:2020dfl,Falkowski:2023hsg,Aoude:2022imd,Severi:2022qjy,Aoude:2023hxv}. 

The framework of the Standard Model Effective Field Theory (SMEFT) is based on the fundamental observation that the high-energy modes of a theory can be integrated out from the Lagrangian when doing low-energy phenomenology, with their effects included via higher-dimensional operators. Over the years, SMEFT has emerged as a versatile and potent approach, owing to its capacity to encompass a multitude of new physics (NP) models while maintaining consistency as a QFT. However, its broad scope results in a plethora of new degrees of freedom, already at the lowest order in the EFT expansion. Given the abundance of SMEFT operators and the presence of unconstrained ({\it flat}) directions in parameter space, it becomes imperative to combine the largest set of measurements to obtain tight constraints, and eventually pave the way for a discovery. Efforts in this direction are evident in recent studies \cite{Aguilar-Saavedra:2018ksv,Hartland:2019bjb,Bissmann:2019gfc,Brivio:2019ius, Banelli:2020iau, Ellis:2020unq,Bissmann:2020mfi,Ethier:2021bye, Miralles:2021dyw, Liu:2022vgo, Kassabov:2023hbm, Allwicher:2023shc, Elmer:2023wtr,Celada:2024mcf}. In the endeavor to maximize the experimental information obtainable from data, one is naturally driven to the consideration of new and exotic observables. In this regard, quantum measurements have been found to be particularly promising \cite{Maltoni:2024tul,Aoude:2022imd,Severi:2022qjy,Fabbrichesi:2022ovb,Fabbrichesi:2024xtq,Aoude:2023hxv,Fabbrichesi:2023jep}, motivating their inclusion in global EFT interpretations. 

Whilst quantum observables have been extensively studied for top production at the LHC, the prospects of measuring them at future lepton colliders remain unexplored. The nature of the initial state, the fact that the dominant production mode is through EW interactions, and the clean environment of a lepton collider, make this a particularly interesting case study.  In this work we set the stage by exploring for the first time the phenomenology of quantum spin observables in top pair production, including spin correlations, entanglement, and violations of Bell inequalities. We also study the impact of quantum observables in constraining heavy new physics in the top sector within the SMEFT framework. 

The work is organised as follows: in Section \ref{sec:basics} we will briefly recapitulate the basics of spin measurements at colliders underlying the difference between hadron-hadron and lepton-lepton machines. The spin correlations in SM electroweak top-pair production are analysed in Sections \ref{sec:sm_ew},~\ref{sec:collphen} and \ref{sec:parity}, where we present analytical results for the correlation matrix and discuss various aspects of lepton collider phenomenology in the SM. Next we move on to SMEFT, with analytical results in Section \ref{sec:eft} and a simulated analysis for FCC-$ee$ and other future colliders in Section \ref{sec:analysis}.

\section{Top spin correlations and entanglement}\label{sec:basics}

\begin{figure}[H]
\centering
 \includegraphics[width=0.30\textwidth]{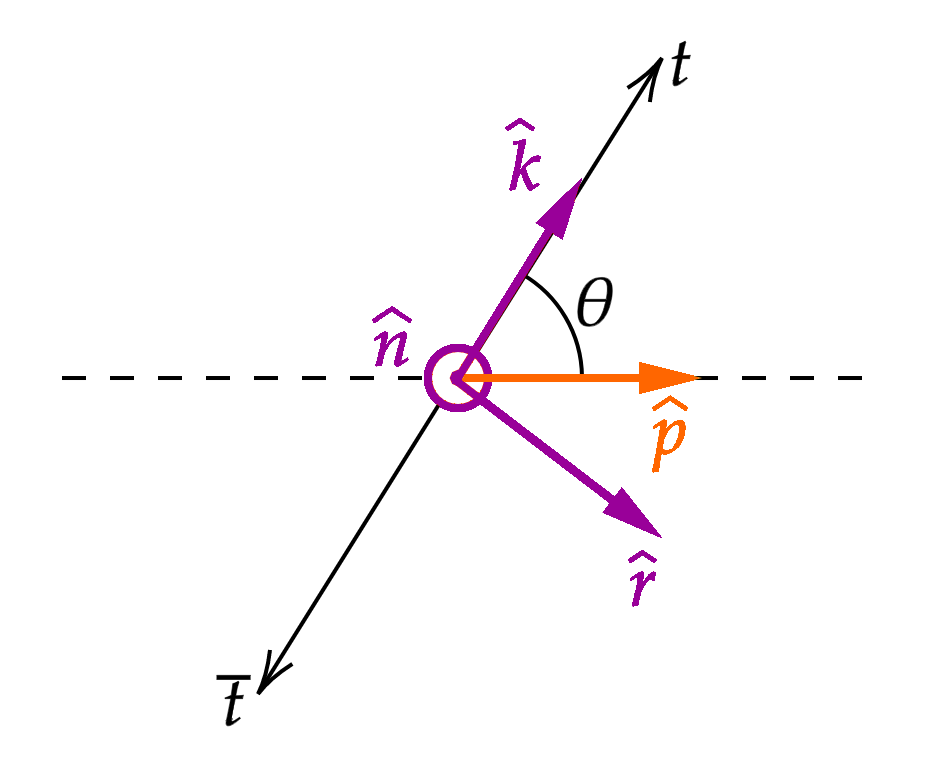}
  \caption{Basis used to characterise the spin state of a $t \bar t$ pair, drawn in the top pair reference frame. The positive beam direction $\hat p$ is the direction of flight of the positive lepton.}
 \label{fig:theta}
\end{figure}

Spin correlations of top quarks, like all two-qubit systems, require nine degrees of freedom to be described. The usual parameterisation is to write the spin density matrix as:
\begin{equation}
    \rho = \nicefrac{1}{4} \big( \mathbf{1} \otimes \mathbf{1} + \mathcal B_{1} \cdot  \boldsymbol{\sigma} \otimes \mathbf{1} + \mathcal B_{2} \cdot  \mathbf{1} \otimes \boldsymbol{\sigma}  + \mathcal C \cdot \boldsymbol{\sigma} \otimes \boldsymbol{\sigma} \big), \label{rho}
\end{equation}
where $\mathcal B_1 = \lbrace B_{1i} \rbrace$ and $\mathcal B_2 = \lbrace B_{2j} \rbrace$ describe the polarisation of the top and antitop quark around the $i$-th and $j$-th axes respectively, and $\mathcal C = \lbrace C_{ij} \rbrace$ represents the amount of correlation between the $i$-th component of the top spin and the $j$-th component of the anti-top spin.  To determine the entries of $\mathcal B$ and $\mathcal C$ a basis in space must be chosen. As it is typically done, in this work we will use a basis constructed from the top quarks direction of flight,
	\begin{equation}
		\hat k = \text{top direction}, \quad	\hat r = \frac{ \hat p - \hat k \, \cos \theta}{\sin \theta}, \quad \hat n = \frac{\hat p \times \hat k}{\sin \theta}	 \label{helbasis} \,,
	\end{equation}
where $\hat p$ represents the $e^+$ beam direction, and $\theta$ is the scattering angle of the top quark with respect to $\hat p$ in the $t \bar t$ rest frame, $\cos \theta = \hat k \cdot \hat p$, see Figure \ref{fig:theta}.

For instance, using the degrees of freedom in \eqref{rho}, the pure spin-singlet state and the three pure spin-triplet Bell states are given by:
\begin{equation}
    \mathcal C^{(\text{singlet})} = 
    \begin{pmatrix}
     -1 & 0 & 0 \\
     0 & -1 & 0 \\
     0 & 0 & -1
    \end{pmatrix}, \quad  \mathcal C^{(\text{triplet})} = 
    \Big \lbrace \begin{pmatrix}
     -1 & 0 & 0 \\
     0 & 1 & 0 \\
     0 & 0 & 1
    \end{pmatrix}, \
    \begin{pmatrix}
     1 & 0 & 0 \\
     0 & -1 & 0 \\
     0 & 0 & 1
    \end{pmatrix}, \
    \begin{pmatrix}
     1 & 0 & 0 \\
     0 & 1 & 0 \\
     0 & 0 & -1
    \end{pmatrix}\Big \rbrace. \label{four}
\end{equation}

Determining the presence of entanglement in a two-qubit system, such as the $t \bar t$ pair's spins, is a solved problem. A generic measure of entanglement for a bipartite system is the so-called entanglement of formation $E_{F}$, which measures how much entanglement is necessary on average to prepare the state (measured in ebit, i.e., the entanglement of a Bell pair).  For a two-qubit system it can be expressed as~\cite{Wootters:1997id}:
\begin{equation}
    E_{F}(\rho) = H\Big(\frac{1+\sqrt{1-C^{2}}}{2}\Big), \label{EFdef}
\end{equation}
where $H$ is Shannon's entropy function,
\begin{equation}
    H(x) = -x \log_2 x - (1-x) \log_2 (1-x).
\end{equation}
The quantity $C$ appearing in \eqref{EFdef} is the two-qubit concurrence, computed explicitly as:
\begin{equation}
    C = \max (0,\lambda_{1}-\lambda_{2}-\lambda_{3}-\lambda_{4})\,,\label{eq:Cvslambdas}
\end{equation}
 in which $\lambda_{1}\ge\lambda_{2}\ge\lambda_{3}\ge\lambda_{4}$ are the eigenvalues of the $4\times4$ matrix: 
 \begin{equation}
     \sqrt{ \sqrt{\rho} \, (\sigma_y \otimes \sigma_y) \rho^\star (\sigma_y \otimes \sigma_y) \, \sqrt{\rho}}, \label{forconcurrence}
 \end{equation}
 and $\rho^\star$ is the complex-conjugated density matrix expressed in the computational basis. We note that:
\begin{equation}
    E_F = 0 \ \iff \  C = 0,
\end{equation}
so non-zero concurrence is equivalent to non-zero entanglement of formation, and since \eqref{EFdef} is a monotonically increasing function of $C$, more concurrence is equivalent to more entanglement.

Unfortunately, the concurrence, and therefore entanglement of formation, are likely to be problematic observables for collider experiments. This is due to the procedure one has to follow to extract these  quantities from real data that tends to systematically bias any initial (however small) experimental uncertainty upwards. Even more worryingly, similarly defined observables are also known to exhibit an upward-bias in their central value \cite{Severi:2021cnj}. However, as it is well known \cite{2203.05582, 2205.00542, Maltoni:2024tul}, for the purposes of establishing/measuring the entanglement of spin, it is sufficient to consider the four markers:
\begin{align}
   D^{(1)} &= \nicefrac 1 3 (+ C_{kk} + C_{rr} + C_{nn}), \label{D1} \\
    D^{(k)} &= \nicefrac 1 3 (+ C_{kk} - C_{rr} - C_{nn} ), \\
    D^{(r)} &= \nicefrac 1 3 ( - C_{kk} + C_{rr} - C_{nn} ), \\
    D^{(n)} &= \nicefrac 1 3 (- C_{kk} - C_{rr} + C_{nn} ).\label{D4}
\end{align}
A sufficient condition for the presence of entanglement in \eqref{rho} is that:
\begin{equation}
    D_{\min} \equiv \min \lbrace D^{(1)}, D^{(k)}, D^{(r)}, D^{(n)}\rbrace, \label{dmin}
\end{equation}
satisfies:
\begin{equation}
    D_{\min} < - \nicefrac 1 3 \qquad \text{(entanglement)}. \label{Dentanglement}
\end{equation}
Indeed, the first observations of entanglement between top quarks \cite{ATLAS:2023fsd, CMS:2024pts} have been obtained by establishing that in $pp$ collisions the marker $D^{(1)}$ is smaller than $-\nicefrac 1 3$ in the threshold region ($D^{(1)}$ is the smallest $D$ in this case). 

The condition \eqref{Dentanglement} is in general only {\it sufficient}, but in the special case where the individual top polarisations are zero, $\mathcal B_1 = \mathcal B_2 = 0$, and the spin correlation matrix is diagonal, then \eqref{Dentanglement} is also {\it necessary}, since the concurrence is given by:
\begin{equation}
    C = \frac 1 2 \max\big(0, -1 - 3 D_{\min} \big),
\end{equation}
a relation that clearly shows that non-zero concurrence is obtained for $D_{\min} < -\nicefrac 1 3$.

\begin{figure}[H]
    \centering
    \includegraphics[width=.46\textwidth]{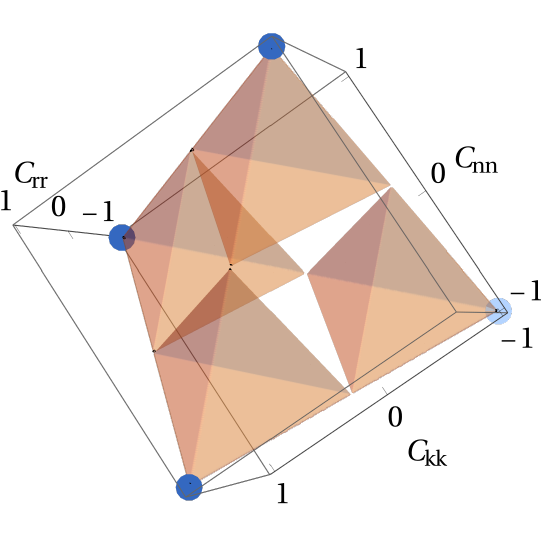}
    \caption{Regions of $\lbrace C_{kk}, C_{rr}, C_{nn} \rbrace$ phase space where the corresponding quantum state is entangled, highlighted in orange. The four pure and maximally entangled states are represented by the blue dots, light blue: singlet, dark blue: triplets. Visualisation based on Ref. \cite{Severi:2021cnj}. }
    \label{fig:ent}
\end{figure}

In Figure \ref{fig:ent} we show the degrees of freedom in the diagonal of $\mathcal C$, highlighting the regions where the corresponding quantum state is entangled, and the four states of \eqref{four}. 

We note that entanglement resides in four disjoint regions, each in the shape of a tetrahedron, inscribed in a tetrahedron of double the edge, and with a pure state at its vertex, while the central cubic region centered on $(0,0,0)$ is not entangled.

\section{Spin correlations in the electroweak SM} \label{sec:sm_ew}

\begin{figure}[t]
    \centering
    \includegraphics[width=.65\textwidth]{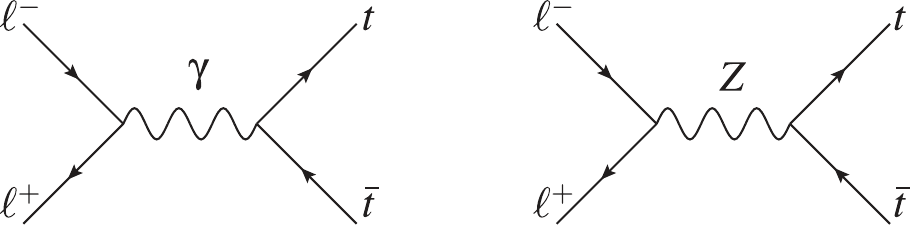}    
    \caption{Feynman diagrams corresponding to tree-level $t \bar t$ production from a lepton pair.}
    \label{fig:tt_SM}
\end{figure}

Top pair production in a lepton collider takes place at LO purely in the $s$--channel, via the exchange of a virtual photon or $Z$ boson, see Figure \ref{fig:tt_SM}. The spin state reached by an $s$-channel vector resonance in $pp$ collisions has been extracted in \cite{Maltoni:2024tul}. Here we show the corresponding results for $\ell^+ \ell^-$ collisions.

Following the notation of \cite{Maltoni:2024tul}, we write the electroweak vertices between two tops and an electroweak vector boson as
\begin{equation}
    \left( t \, \bar t \, A^\mu \right) = i e \Qt \, \gamma^\mu, \qquad \left( t \, \bar t \, Z^\mu \right) = \frac{i e}{\cW \sW}  \, \gamma^\mu \, \left( \gVt \, \mathbbm{1} - \gAt \, \gamma^5 \right) \,, \label{topEW}
\end{equation}
where $g_\text{V} = T_3/2 - Q \, \sW^2$ and $g_\text{A} = T_3/2$, with $T_3 = \nicefrac 1 2$ and $Q = \nicefrac 2 3$ for top quarks.

Following the literature \cite{2003.02280, Aoude:2022imd, Maltoni:2024tul}, we will parameterize the spin correlation matrix $\mathcal C$ as a ratio between an angular-dependent cross section $\widetilde C_{ij}$ and the total rate $A$:
\begin{equation}
C_{ij} = \frac{\widetilde{C}_{ij}}{A}, \label{cij}
\end{equation}
to highlight that, both experimentally and in theoretical calculations, spin observables are extracted as ratios of cross-sections, similarly to, e.g., asymmetries. In case several non--interfering channels contribute to the same process, the total $A$ and $\widetilde C_{ij}$ is simply given by the sum over channels. Further details on the combination of different channels are given in \cite{Maltoni:2024tul}.

Since in a purely $s$-channel process the spin state only depends on the tensor structure inserted in the top and antitop fermion lines, we split the squared matrix element in three parts,
\begin{equation}
    A = A^{[0]} + A^{[1]} + A^{[2]}, \label{a_SM}
\end{equation}
counting the number of $\gamma^5$ insertions in the top quark line. Explicitly, terms in $A$ free of any $\gamma^5$ inserted in the $t \bar t$ fermion line are collected in $A^{[0]}$ (that therefore contains terms with $\Qt^2$, $\gVt^2$, or $\Qt \, \gVt$), terms in $A$ with a single $\gamma^5$ insertion (at orders $\gAt \Qt$ and $\gAt \gVt$) are collected in $A^{[1]}$, and terms with two $\gamma^5$, i.e.\@ terms proportional to $\gAt^2$, are collected in $A^{[2]}$.

The $A^{[0]}$ channel produces the same spin quantum state as QCD in the $q \bar q \to t \bar t$ and $g_L g_R \to t \bar t$ channel, whose properties are well known \cite{2003.02280}, while the spin states reached by the other two channels, $A^{[1]}$ and $A^{[2]}$, are typical of electroweak production. 

Note that there are only two degrees of freedom in the $t \bar t$ kinematics: $\theta$, that is used to construct the helicity basis, and the top velocity $\beta$,
\begin{equation}
    \beta = \sqrt{1 - 4 m_t^2/m_{t \bar t}^2}.
\end{equation}
As a function of $\beta$ and $\sin \theta$ ($\equiv s_\theta$) and $\cos \theta$ ($\equiv c_\theta)$, the spin correlation coefficients are given by: 

\begin{equation}
\begin{dcases}
    A^{[0]} &= F^{[0]}\left( \beta^2 c_{\theta }^2-\beta^2+2  \right), \\
\widetilde{C}^{[0]}_{kk} &= F^{[0]} \left[ \beta^2-\left(\beta^2-2\right) c_{\theta }^2  \right], \\
\widetilde{C}^{[0]}_{kr} &= F^{[0]} \, 2 \sqrt{1-\beta^2} c_{\theta } s_{\theta } , \\
\widetilde{C}^{[0]}_{rr} &= F^{[0]} \, s_{\theta }^2 (2 -\beta^2), \\
\widetilde{C}^{[0]}_{nn} &=  - F^{[0]}  \, \beta^2 s_{\theta }^2,\,
\end{dcases}
\quad
\begin{dcases}
A^{[ 1]} = 2 \, F^{[ 1]} \, c_\theta,  \\
\widetilde{C}_{kk}^{[ 1]} = 2 \, F^{[ 1]} \, c_\theta, \\
\widetilde{C}_{kr}^{[ 1]} = F^{[ 1]} \, \sqrt{1-\beta^2} \, s_\theta, \\
\widetilde{C}_{rr}^{[ 1]} = 0, \\
\widetilde{C}_{nn}^{[ 1]} = 0, 
\end{dcases} \quad
\begin{dcases}
A^{[ 2]} = F^{[ 2]} \, \left(1 + c_\theta^2   \right), \\
\widetilde{C}_{kk}^{[ 2]} = F^{[ 2]} \, \left( 1 + c_\theta^2   \right), \\
\widetilde{C}_{kr}^{[ 2]} = 0, \\
\widetilde{C}_{rr}^{[ 2]} = - F^{[ 2]} \, s_\theta^2, \\
\widetilde{C}_{nn}^{[ 2]} = F^{[ 2]} \, s_\theta^2\,. 
\end{dcases} \label{EW_top_spin}
\end{equation}

The factors $F^{[ 0]}$, $F^{[ 1]}$, and $F^{[ 2]}$, are common to all spin coefficients, and simply parameterize the relative weight of each of the $A^{[i]}$, with $=0,1,2$ channels in \eqref{a_SM}. Throughout this work, we will collect the common factors $F$ in Appendix \ref{app:F}. The factors $F^{[ 0]}$, $F^{[ 1]}$, $F^{[ 2]}$ appear in Eqs.~\@\eqref{F0},\eqref{F1},\eqref{F2}. While in principle higher order corrections to \eqref{EW_top_spin} may be considered to improve accuracy, it is well known that spin observables, being defined as ratios of cross-sections, are largely stable under perturbative corrections, in the SM \cite{Czakon:2020qbd}, as well as the SMEFT \cite{Severi:2022qjy}.

\subsection{Optimal axes}

The properties of the quantum state described by \eqref{EW_top_spin} have been studied for a long time. It is interesting to note that the purely vector $A^{[0]}$ channel produces a $\mathcal C$ matrix with an eigenvalue equal to $1$. This means there exists an {\it optimal axis} $\hat x^{[0]}$ along which spin correlations are maximal:
\begin{equation}
    C_{x x} = 1,
\end{equation}
i.e.\@, the top and antitop spins are aligned with $100\%$ probability. For the quantum state given by $A^{[0]}$, the optimal axis is in the $\hat k-\hat r$ plane and an angle $\xi^{[0]}$ away from the top direction $k$, with:
\begin{equation}
    \tan \xi^{[0]} = \sqrt{1-\beta^2} \, \tan \theta. \label{optvector}
\end{equation}
This axis corresponds to the so-called off-diagonal basis \cite{Mahlon:1997uc} used at Tevatron to maximise spin correlations in $t\bar t$ QCD production, where $q\bar q$ production is dominant. 

The purely axial channel $A^{[2]}$ and the vector--axial interference $A^{[1]}$, taken by themselves without $A^{[0]}$, would also have an optimal axis $\hat x^{[1,2]}$  (different from $x^{[0]}$) such that $C_{x x} = 1$. The optimal axis for $A^{[1]}+A^{[2]}$ is, in fact, the helicity direction:
\begin{equation}
    \tan \xi^{[1,2]} = 0.
\end{equation}
The misalignment between $\hat x^{[0]}$ and $\hat x^{[1,2]}$ in general prevents the full electroweak quantum state, given by $A^{[0]}+A^{[1]}+A^{[2]}$, from having an optimal axis. However, separating top pair production by initial state helicity,
\begin{equation}
    \ell^+_R \, \ell^-_L \to t \bar t, \quad \text{and} \quad \ell^+_L \, \ell^-_R \to t \bar t,
\end{equation}
 one finds that the corresponding full electroweak quantum states both have an optimal axis. The two optimal axes $\hat x^{[\text{RL}]}$ and $\hat x^{[\text{LR}]}$ both lie in the $\hat k-\hat r$ plane, separated by $\hat k$ by the angles:
 \begin{align}
     \tan \xi^{[\text{RL}]} &= \frac{s_\theta \, \sqrt{1 - \beta^2} \, (5 \mt^2 - 2 \cW^2 \mz^2 (1-\beta^2)  \, )}{\mt^2 (5 c_\theta + 3 \beta) - 2 c_\theta \cW^2 \mz^2 (1 - \beta^2)}, \label{optEW1}\\
     \tan \xi^{[\text{LR}]} &= \frac{s_\theta \,\sqrt{1 - \beta^2} ( (2 \cW^2 - 5) \mt^2 + 4 \cW^2 \sW^2 \mz^2 ( 1 - \beta^2 ) \, ) }{\mt^2 ( (2 \cW^2 - 5)  c_\theta + (3 - 6 \cW^2) \beta) + 4 c_\theta \cW^2 \sW^2 \mz^2  (1-\beta^2) }. \label{optEW2}
 \end{align}
The two optimal directions \eqref{optEW1}-\eqref{optEW2} for electroweak top production were first found in \cite{Parke:1996pr}, and our results agree with those therein. 

In practice, the angles $\xi^{[\text{RL}]}$ and $\xi^{[\text{LR}]}$ are numerically very similar. Therefore, the quantum state that is observed in unpolarised lepton colliders, given by the helicity-summed process $\ell^+ \ell^- \to t \bar t$, has an {\it approximate} optimal axis $\hat x$, with $C_{xx}$ less than, but very close to, $1$. This observation, already noted in \cite{Parke:1996pr}, holds true across the $t \bar t$ phase space. For instance, the spin correlation obtained for $\beta = 0.33$, \footnote{Corresponding to $m_{t \bar t} = 365 \, \text{GeV}$, one of the proposed operating energies of FCC-$ee$.} and $\theta = \pi/2$ on the optimal axis is:
\begin{equation}
    C_{xx} = 0.9996.
\end{equation}

\section{Collider phenomenology within the electroweak SM} \label{sec:collphen}

\begin{figure}[h]
    \centering\includegraphics[width=.46\textwidth]{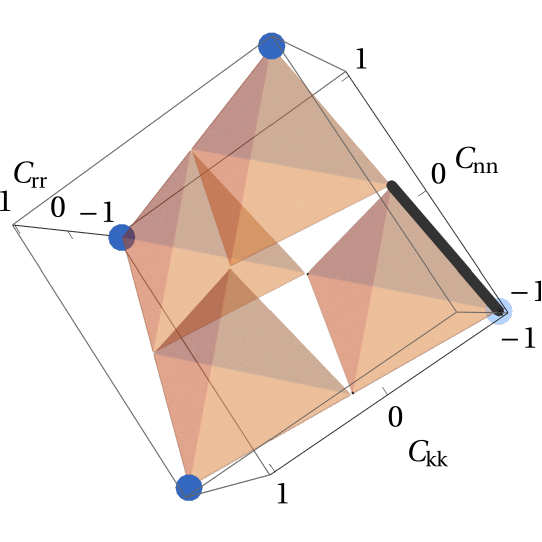}     \includegraphics[width=.46\textwidth]{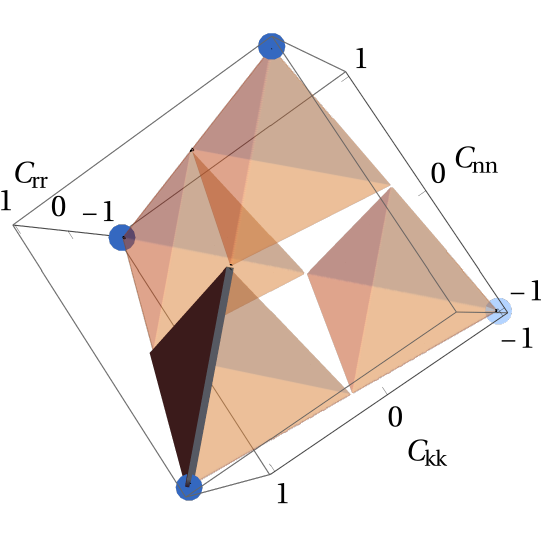}
    \caption{Ensemble of quantum states, in black, reachable by QCD top pair production in a hadron collider from the initial states $g_R \, g_R$ and $g_L \, g_L$, left, and $g_L \, g_R$ and $q \bar q$ (any helicity), right, overlaid on Figure \ref{fig:ent}. The plotted surface contains all spin quantum states obtained varying the partonic energy $\sqrt{\hat s}$ and the top scattering angle $\theta$ within their physically allowed ranges.}
    \label{fig:ent_protons}
\end{figure}

We now compare the properties of the $t \bar t$ spin quantum state reached from proton--proton collisions (e.g.\@ at the LHC) with the one obtained from lepton--lepton collisions of various energies. 
The spin quantum state reached in proton--proton colliders is well known, and we refer to dedicated publications \cite{2003.02280, Fabbrichesi:2021npl, Severi:2021cnj, 2203.05582, 2205.00542, 2209.03969} for further details. For the purposes of comparing the reach of a $pp$ machine with that of a lepton collider, in Figure \ref{fig:ent_protons} we show the quantum states produced by QCD--mediated top pair production in the $gg$ and $q \bar q$ channels. 
For convenience, in the plot we have split the $gg$ initial state by helicity, because of the difference in the respective $t \bar t$ spin states. No such split is needed for $q \bar q$, since all initial helicity configurations result in the same $t \bar t$ spin. 

We recall that in a $pp$ machine there are two regions of interest for quantum information studies in $t \bar t$: the {\it threshold} region, given by:
\begin{equation}
    \beta \approx 0,
\end{equation}
corresponding to top quarks produced (almost) at rest, and the {\it boosted} region, given by:
\begin{equation}
    \beta \approx 1, \quad \cos \theta \approx 0,
\end{equation}
corresponding to top quarks produced at large velocity, and almost perpendicular to the beam. Once all channels are combined with the relative PDFs, at threshold $t \bar t$ production from gluon fusion dominates, and produces an entangled, singlet state. Experimental studies \cite{ATLAS:2023fsd,CMS:2024pts} have defined this {\it threshold} region to be approximately given by:
\begin{equation}
    \beta < 0.42 \quad \iff \quad m_{t \bar t} < 380 \, \text{GeV} \label{lhc_thresh}
\end{equation}
On the contrary, at large top $p_T$ the state tends to a pure triplet regardless of the initial state. This {\it boosted} region can be approximately defined as \cite{Severi:2021cnj,Han:2023fci,CMS:2024vqh}:
\begin{equation}
    \begin{dcases}
        \beta > 0.90 \quad \iff \quad m_{t \bar t} > 800 \, \text{GeV} \\
        \cos \theta < 0.2 \quad \iff \quad 79^\circ < \theta < 101^\circ.
    \end{dcases}
     \label{lhc_boosted}
\end{equation}
Finally, the intermediate transition region, at $p_T \sim \mt$, is not entangled. 

\begin{figure}[H]
    \centering
    \includegraphics[width=.46\textwidth]{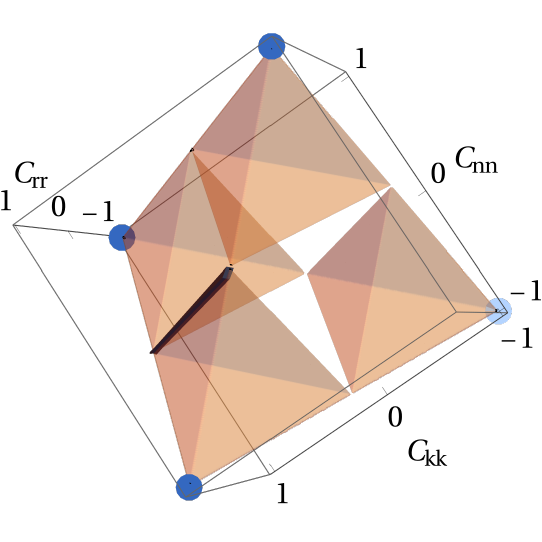}
    \includegraphics[width=.46\textwidth]{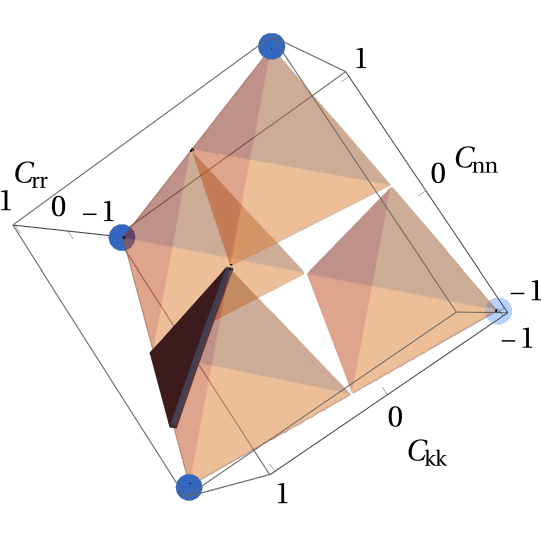}
    \caption{Ensemble of quantum states, in black, reachable by a lepton collider operating at $\sqrt{s} \sim 2 \mt$, left, and at $\sqrt{s} \gg 2 \mt$, right, overlaid on the entangled regions of Figure \ref{fig:ent}. Note that the black region lies on the side of the bottom--left pyramid, not inside it.
    }
    \label{fig:ent_lep}
\end{figure}

Moving to lepton colliders, similarly to Figure \ref{fig:ent_protons}, in Figure \ref{fig:ent_lep} we plot the ensemble of quantum states reachable in a lepton collider operating near $t \bar t$ threshold and at $\sqrt{s} \gg \mt$.

\begin{figure}
    \centering
    \includegraphics[width=.9\textwidth]{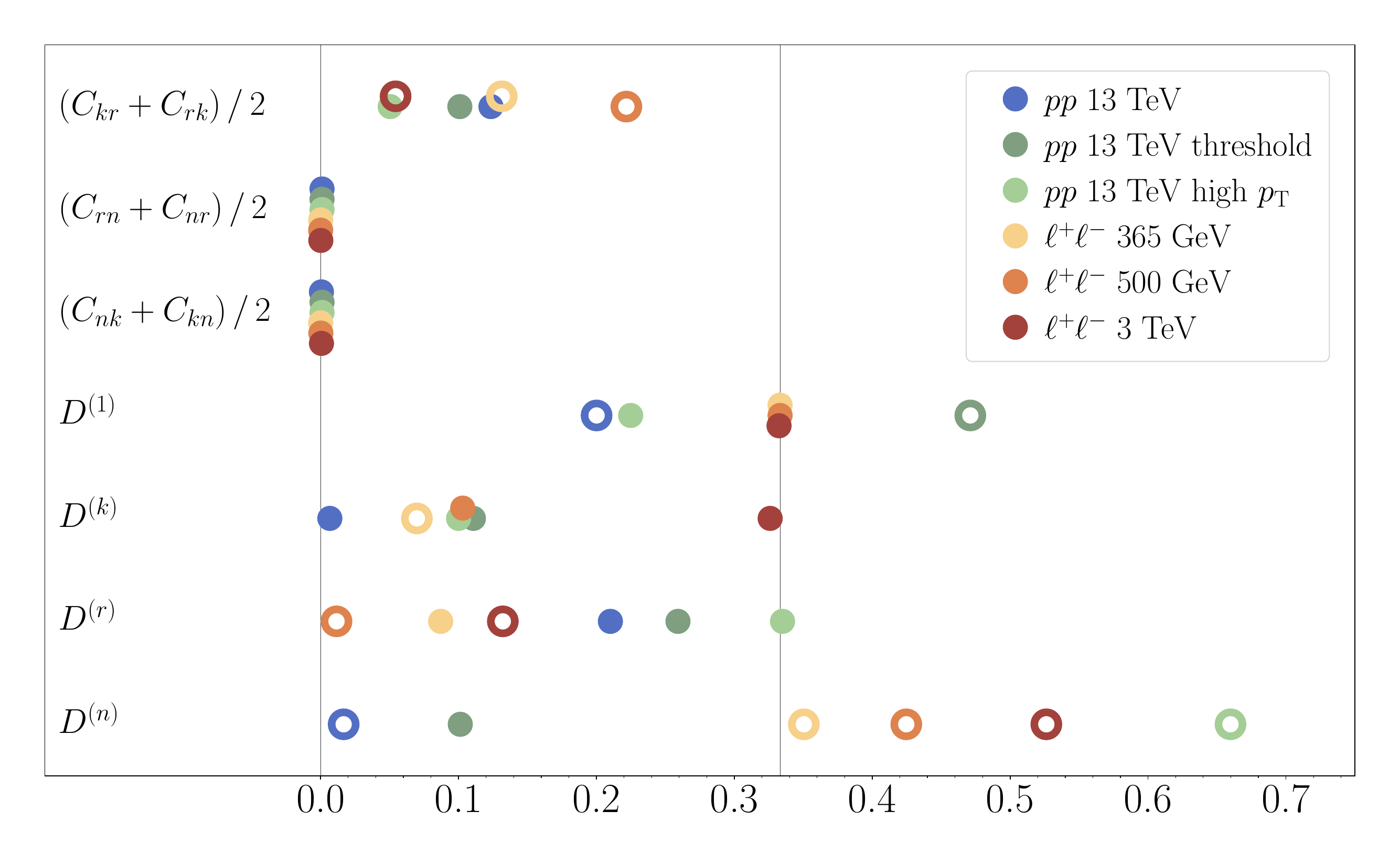}
    \caption{Predicted values of the $t \bar t$ top spin correlation and entanglement markers $D$ and $C_{ij} + C_{ji}$ at parton-level, for various lepton collider energies and for the LHC.  We plot the absolute values of the markers: full circles identify a positive value, while empty ones indicate a negative value. Overlapping markers are spread vertically for readability. We remind the reader that entanglement is reached when empty points appear on the right of the vertical line at $\nicefrac 1 3$.}
    \label{fig:spin_sm}
\end{figure}

To quantitatively compare the different properties of the $t \bar t$ spin states at the LHC and at future lepton colliders, in Figure \ref{fig:spin_sm} we show our parton--level predictions for a top/anti-top spin correlation measurement in several proposed future lepton colliders. The results are based on a simulation with {\tt Madgraph5\_aMC@NLO} accurate at leading-order in the strong and electroweak couplings, and using the {\tt NNPDF4.0} \cite{NNPDF:2021njg} parton distribution function for the LHC. It is interesting to note that due to the particular spin structure produced by the electroweak sector of the SM, the value of $D^{(1)}$ in lepton collisions is fixed\footnote{Note the $+$ sign.}:
\begin{equation}
    \nicefrac 1 3 \, \text{Tr}\left[ \mathcal C \right] = D^{(1)} = + \, \frac 1 3, \label{DfixSM}
\end{equation}
for all production channels in \eqref{EW_top_spin}, for any collider energy and for any $t \bar t$ kinematics, i.e.\@ for any scattering angle. 
The relation \eqref{DfixSM} is a general consequence of the spin--triplet state reached by the exchange of an $s$-channel spin-one boson, and it is also valid at the LHC for the $g_L g_R$ and $q_R \bar q_L = q_L \bar q_R$ channels, but not for the (dominant) $g_R g_R$ and $g_L g_L$ ones. The relation \eqref{DfixSM} is broken by higher order corrections, such as the emission of an extra jet, that randomises and dilutes spin correlations, resulting in a smaller value of $D^{(1)}$.

We also find that two of the three off-diagonal spin correlations, $C_{rn} + C_{nr}$ and $C_{kn} + C_{nk}$, are consistent with zero both in proton and in lepton collisions at any energy, due to the absence, at LO, of $P$-violating absorptive contributions to the $t \bar t$ production amplitude \cite{Bernreuther:2015yna}. The third off-diagonal correlation $C_{kr}+C_{rk}$ is larger than at the LHC, and generally decreasing with energy.

\subsection{Entanglement}

\begin{figure}[h]
    \centering
        \begin{minipage}{0.54\textwidth}
        \centering
        \includegraphics[width=\textwidth]{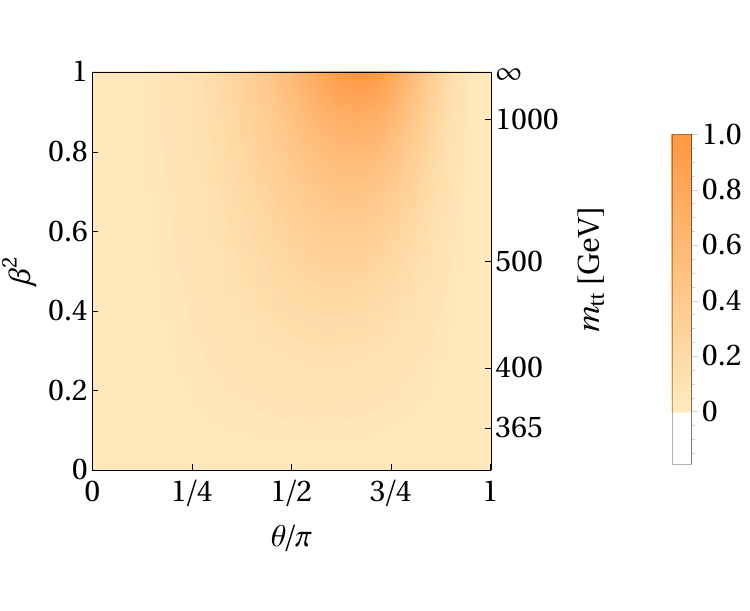}
    \end{minipage}\hfill
    \begin{minipage}{0.43\textwidth}
        \centering
        \includegraphics[width=\textwidth]{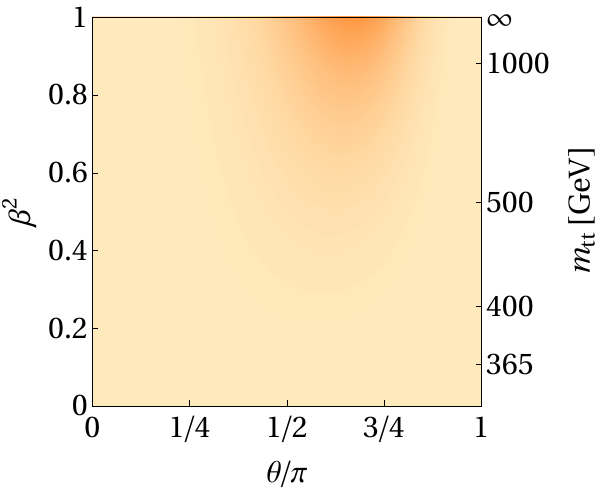}
    \end{minipage}
    \vspace{-5mm}
    \caption{Value of the concurrence, left, and of the entanglement of formation, right, for the $t \bar t$ quantum state reached by a lepton collision, as a function of the top kinematics $\beta$ (or $m_{t \bar t}$) and $\theta$.}
    \label{fig:concurrence}
\end{figure}

We can show explicitly the entanglement structure in $t \bar t$ phase space for lepton colliders by plotting the concurrence \eqref{eq:Cvslambdas} and the entanglement of formation \eqref{EFdef}, in Figure \ref{fig:concurrence}.

Unlike the LHC, in a lepton collider concurrence/entanglement is present in all $t \bar t$ phase space. Both for the LHC and for circular lepton colliders, the point of maximum concurrence/entanglement corresponds to top quarks being produced ultra-relativistically and in a (calculable) optimal direction, 
\begin{equation}
    \beta = 1, \quad \theta = \theta_{\max}\,.
\end{equation}
 At the LHC the parity-conserving nature of strong interactions implies $\theta_{\max} = \pi/2$, while in a lepton collider the optimal angle is $\theta_{\max} > \pi/2$, that is, a configuration where the positively charged top quark is closer to the positively charged lepton\footnote{In a $pp$ collider the beam is symmetric, and there is no opportunity for such asymmetry.}. 

 Moving to observables more relevant to collider phenomenology, we find that the smallest $D$ is always $D_{\min} = D^{(n)}$, for which we obtain: 
\begin{align}
    D^{(n)}_{\sqrt s \, = \, \text{365 GeV}} &= -0.35, \\
    D^{(n)}_{\sqrt s \, = \, \text{500 GeV}} &= -0.42, \\
    D^{(n)}_{\sqrt s  \, = \, \text{3 TeV}} &= -0.53.\end{align}

While the classical bound of $-\nicefrac 1 3$ is always violated, at least at parton level, the size of $D$ in all cases is smaller than what is available at the LHC. 
For comparison, in the $t \bar t$ threshold region \eqref{lhc_thresh} where entanglement between top quarks has been observed for the fist time \cite{ATLAS:2023fsd}, we obtain:
\begin{equation}
    D^{(1)}_{\text{LHC thr}} = -0.47,
\end{equation}
 while in the boosted region \eqref{lhc_boosted}, we obtain:
\begin{equation}
    D^{(n)}_{\text{LHC boost}} = -0.66.
\end{equation}

\subsection{Bell violations}

We now briefly discuss the possibility of observing a violation of Bell inequalities in a lepton collider. Bell violations are the ultimate test for ``quantumness'' in a physical system, as they require the existence of correlations so strong that no classical model, no matter how concealed or intricate, can explain.  Violations of Bell inequalities have been observed in a variety of physical systems, but never at the hundreds of GeV/several TeV scale that is accessible in modern particle colliders. Recently, several proposals \cite{Fabbrichesi:2021npl, Severi:2021cnj, Dong:2023xiw, Han:2023fci} have been put forth for the tentative detection of Bell violations with top quark pairs at the LHC. A violation of Bell inequalities (in the CHSH form) in two-state systems such as a $t \bar t$ pair is given by the expectation value of the Bell operator $\mathcal B$ exceeding $2$:
\begin{equation}
    \langle a \, b + a \, b' + a' \, b - a' \, b'  \rangle \equiv \langle \mathcal \, {\mathcal B}(a,a',b,b') \, \rangle > 2, \quad \implies \quad \text{Bell violation}.
\end{equation}
We have denoted by $a^{(}{'}^{)} \, b^{(}{'}^{)}$ the outcome of a measurement of the first particle along axis $\hat a^{(}{'}^{)}$ and of the second particle along axis $ \hat b^{(}{'}^{)}$. Given a (mixed) quantum state of two qubits, determining the directions $ \hat a, \hat a', \hat b, \hat b'$ for which the Bell operator is maximised is a solved problem. At the LHC, the optimal axes have been found in \cite{Severi:2021cnj}.  Incidentally, we note that a sufficient condition for Bell violations is that the $D_{\min}$ of Eq.~\eqref{dmin} satisfies:
 \begin{equation}
    D_{\min} < - \frac{1}{\sqrt 2} \qquad \text{(Bell violation)}.
\end{equation}

\begin{figure}[t]
    \centering
        \begin{minipage}{0.54\textwidth}
        \centering
        \includegraphics[width=\textwidth]{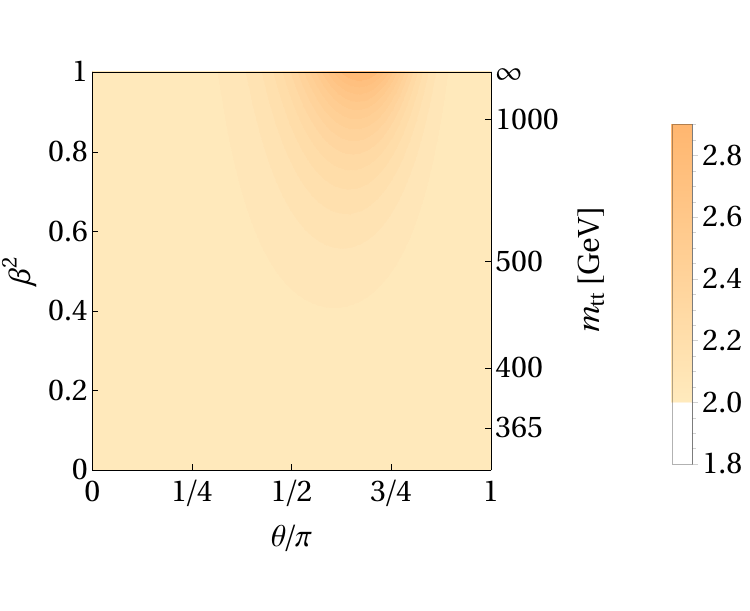}
    \end{minipage}\hfill
    \begin{minipage}{0.43\textwidth}
        \centering
        \includegraphics[width=\textwidth]{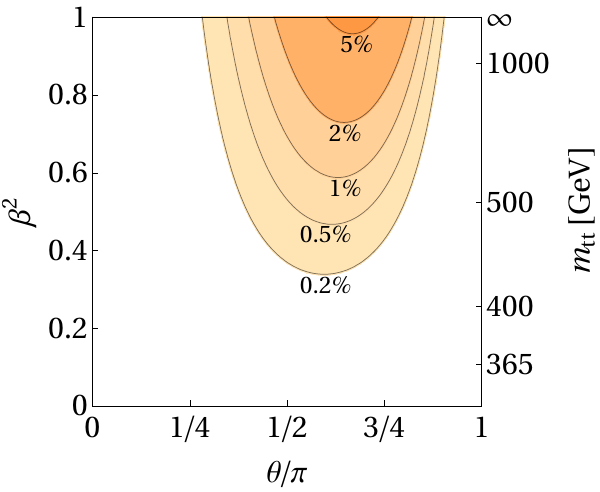}
    \end{minipage}
    \vspace{-5mm}
    \caption{Left: expectation value of the Bell operator $\mathcal B(a,a',b,b')$ evaluated on the optimal directions $ \hat a, \hat a', \hat b, \hat b'$  for the $t \bar t$ quantum state reached by a lepton collision, as a function of the top kinematics $\beta/m_{t \bar t}$ and $\theta$. Right: experimental accuracy needed to establish $\langle \mathcal B \rangle > 2$ at $5 \sigma$ using events at any particular $(\beta, \theta)$.}
    \label{fig:bell}
\end{figure}

In general, Bell inequalities are not expected to be violated in the entirety of $t \bar t$ phase space, but only in a small corner that maximises spin entanglement. In Fig.~\ref{fig:bell} we show the expectation value of $\langle \mathcal B \rangle$ on $t \bar t$ pairs produced from lepton collisions, as a function of $\beta$ and $\theta$, on the optimal directions $\hat a, \hat a', \hat b, \hat b'$. Furthermore, to directly assess the feasibility of a Bell violation experiment in a lepton collider, in Fig.~\ref{fig:bell} we also plot the minimum experimental accuracy on $\langle \mathcal B \rangle$ that would be required to establish a violation at $5 \sigma$. 

Contrary to the LHC case, in a lepton collider a Bell violation is always present in the SM, at least at parton level. The degree of violation $\langle \mathcal B \rangle - 2$ is however very small in all $t \bar t$ phase space, apart from a small region at large $\beta$ and around $\theta_{\max} \gtrsim \pi/2$, that we had already found when considering the concurrence in Figure \ref{fig:concurrence}. As already noted, larger $\sqrt{s}$ (or $\beta$) means more entanglement and therefore a more significant Bell violation.  In fact, a machine operating near $t \bar t$ threshold is unlikely to observe Bell violations at all. If the Bell operator is reconstructed to a total  accuracy of $2\%$, we find that a $750 \, \text{GeV}$ $\ell^+ \ell^-$ collider could establish Bell violations. If instead the final accuracy is closer to $5 \%$, a TeV-scale collider would be needed. 

We finally remark that, even if the precision on $\langle \mathcal B \rangle$ will prove to be significantly better than anticipated here, Bell tests at low $\beta$ suffer from the locality loophole \cite{Severi:2021cnj}, so, from the point of view of quantum information, a test at the TeV scale is preferable in any case.

\subsection{Parametrisation of $t \bar t$ threshold effects}

 Lepton colliders will offer a unique window of exploration for threshold effects in top pair production. It is well known that top quark pairs produced near threshold feel each other's presence, and evolve subject to a QCD potential \cite{Hoang:2001mm, Hoang:2013uda, Bach:2017ggt}, which, applied to stable top quarks, predicts pseudoscalar ${}^{2S+1}L_J^{[\rm col]} = {}^1S_0^{[1]}$ $(\eta)$, and vector ${}^3S_1^{[1]}$ ($\psi$) states to be the most bound, and almost degenerate in mass ($\delta m\lesssim$ 100 MeV). Top quarks generally decay before such $t\bar t$--mesons fully form, but the effect of the binding is still visible in the invariant mass distribution and in other observables.

In \cite{Maltoni:2024tul} we introduced an effective model with a pseudo-scalar resonance $\eta$, with the intent to parameterize the bulk of bound-state effects at the LHC.
In the same spirit, bound state effects in a $\ell^+ \ell^-$ collider can be mimicked by the inclusion of a vector resonance $\psi$, with the approximate parameters:
\begin{equation}
    m_\psi = m_\eta \simeq 2 \mt - 2 \, \text{GeV}, \quad \text{and} \quad \Gamma_\psi = \Gamma_\eta \simeq 2\, \Gamma_{\text t}.
\end{equation}
In our model, implemented in {\tt MadGraph5\_aMC@NLO}, the vector state $\psi$ is produced from $\ell^+ \ell^-$ annihilation and decays into top quarks, with a coupling tuned to reproduce the total cross-section predicted by the QCD potential. In our simulation, we force the $\psi$ to be resonant with a Breit-Wigner shape. The $\psi$ contribution is then added (at the amplitude squared level) to the fixed-order calculation for $W^+ b W^- \bar b$ production in the threshold region, at NLO QCD for the lepton collider and at LO QCD with the NLO $K$-factor for the LHC.

\begin{figure}[h]
    \centering
    \begin{minipage}{.495\textwidth}
      \centering
      \includegraphics[width=\linewidth]{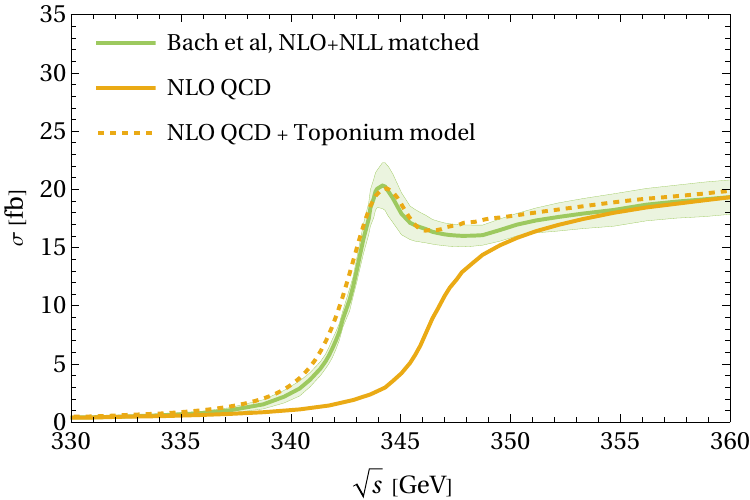}
    \end{minipage}%
    \begin{minipage}{.495\textwidth}
      \centering
      \includegraphics[width=\linewidth]{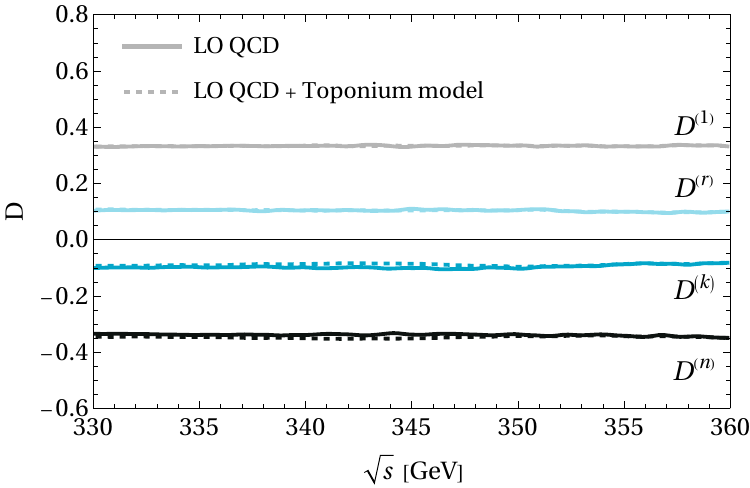}
    \end{minipage}
    \begin{minipage}{.495\textwidth}
      \centering
      \includegraphics[width=\linewidth]{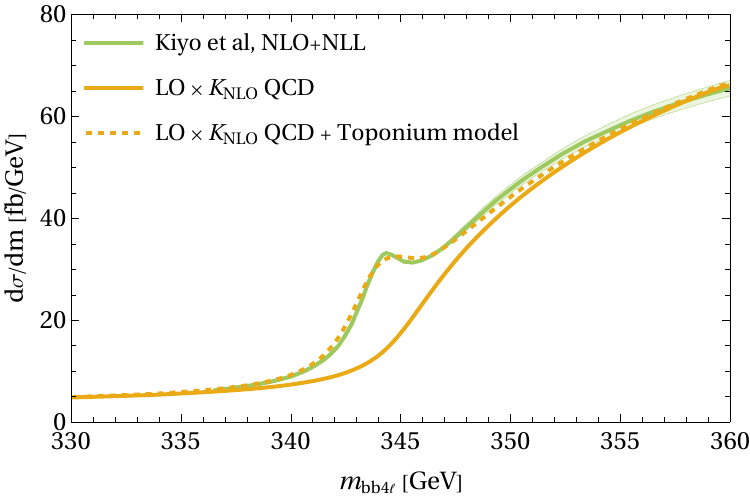}
    \end{minipage}%
    \begin{minipage}{.495\textwidth}
      \centering
      \includegraphics[width=\linewidth]{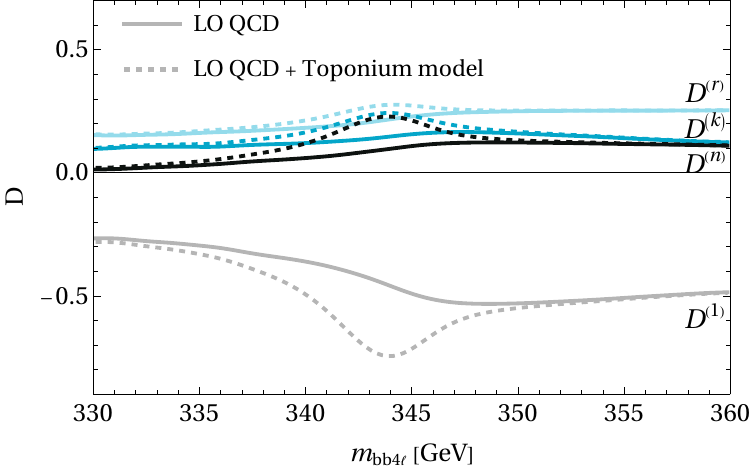}
    \end{minipage}
    \caption{Left: total cross-section for $t \bar t$ production near threshold at fixed-order, with the inclusion of the QCD potential \cite{Bach:2017ggt}, and with our effective model. Right: entanglement markers $D$ at fixed-order and with our model. For reference, the bottom row shows the same plots for the LHC, at fixed order, with the QCD potential \cite{Kiyo:2008bv}, and with our effective model.}
    \label{fig:toponium2}
\end{figure}

 Our results are shown in Fig.~\ref{fig:toponium2}. We find it remarkable that by just fitting one parameter, {\it i.e.,}  the coupling of the resonance to the top quark through the height of the peak, such a simple model can reproduce the resummed analytic results for the threshold cross-section, within the theoretical uncertainty band. The corresponding predictions for the entanglement observables $D$ of \eqref{D1}--\eqref{D4}, are also shown in Fig.~\ref{fig:toponium2}. The signature of formation of a $\psi$ state is a localised increase in the total rate with no significant changes in the spin state, since the vector resonance has the same spin quantum numbers as the photon and $Z$ boson, which already mediate top pair production.  We recall that this is in contrast with the LHC, where the formation of an $\eta$ enhances $D^{(1)}$, as shown in the bottom plots of Fig.~\ref{fig:toponium2}~\cite{Maltoni:2024tul}.

\section{Entanglement and parity symmetry}\label{sec:parity}

It has been noted that several physical phenomena, such as the scattering of Kerr black holes \cite{Aoude:2020mlg}, certain properties of QCD hadrons \cite{Beane:2018oxh}, and even the Higgs \cite{Carena:2023vjc} and electroweak \cite{Cervera-Lierta:2017tdt} sector of the Standard Model, seem to obey criteria of minimisation of entanglement. In this regard, several works \cite{Low:2021ufv,Beane:2018oxh} conjectured  a possible relation between the extremisation of entanglement and the emergence of additional symmetries. 

In this Section we show that $t \bar t$ production in a lepton collider constitutes another example of a physical system where points of extremal entanglement are related to the restoration of symmetries. \smallskip

The electroweak interactions responsible for top pair production \eqref{topEW} conserve parity either when
\begin{equation}
    \gAt = 0, \label{gAt0}
\end{equation}
or 
\begin{equation}
    \Qt = 0 \text{  and  } \gVt = 0. \label{gVt0}
\end{equation}
In the SM one has $\gAt = \nicefrac 1 4$, $\Qt = \nicefrac 2 3$, and $\gVt = \nicefrac 1 4 - \nicefrac 2 3 \, \sW^2$, and in fact electroweak interactions violate parity.

If the top production channel is $P$--conserving and purely vectorial ($\gAt = 0$) the ensemble of quantum states available becomes the same as $q \bar q \to t \bar t$ in QCD, on the right of Figure \ref{fig:ent_protons}. We note in particular that the maximally--entangled and pure triplet state:
\begin{equation}
    \mathcal C = \begin{pmatrix}
     1 & 0 & 0 \\
     0 & 1 & 0 \\
     0 & 0 & -1
    \end{pmatrix}, \label{pure_P}
\end{equation}
becomes available when $\gAt = 0$, for the configuration $\theta = \pi/2$ and $\beta \to 1$. In this limit, we find from \eqref{EW_top_spin}:
\begin{equation}
\begin{dcases}
C^{[0]}_{kk} = 1, \\
C^{[0]}_{kr} = 0 , \\
C^{[0]}_{rr} = 1, \\
C^{[0]}_{nn} =  -1,\,
\end{dcases}
\qquad
\begin{dcases}
C_{kk}^{[ 1]} = 1,\\
C_{kr}^{[ 1]} = 0, \\
C_{rr}^{[ 1]} = 0, \\
C_{nn}^{[ 1]} = 0, 
\end{dcases} \qquad
\begin{dcases}
C_{kk}^{[ 2]} = 1 , \\
C_{kr}^{[ 2]} = 0, \\
C_{rr}^{[ 2]} = -1, \\
C_{nn}^{[ 2]} = 1\,, 
\end{dcases}
\qquad (\theta = \pi/2, \ \beta \to 1). \label{clim}
\end{equation}
While the $P$--conserving channel "$0$" indeed reaches \eqref{pure_P}, {\it any} amount of contribution from the axial couplings induces the $P$-violating interference channel "$1$", and moves the total state away from the maximally entangled point. 

Conversely, if top production is purely axial ($\Qt = \gVt = 0$), it is also $P$--conserving, only the "$2$" channel survives, and the maximally-entangled pure triplet:
\begin{equation}
    \mathcal C = \begin{pmatrix}
     1 & 0 & 0 \\
     0 & -1 & 0 \\
     0 & 0 & 1
    \end{pmatrix}, \label{pure_P1}
\end{equation}
is accessible for $\theta = \pi/2$ and $\beta \to 1$. Again, any pollution from the "$1$" state makes the sum $P$--violating, and spoils purity. \smallskip

To show these effects in action, we consider top pair production mediated by a $Z$ boson, with arbitrary vector and axial couplings:
\begin{equation}
    \mathcal L \supset \, \frac{e}{\sW \cW} \big( \, Z_\mu \, \overline t \, \gamma^\mu ( \gVt + \gamma^5 \gAt) \, t + \, Z_\mu \, \overline \ell \, \gamma^\mu ( \gVl + \gamma^5 \gAl) \, \ell \, \big).
\end{equation}
In the SM we have for the top quark $\gAt = \nicefrac 1 4$ and $\gVt = \nicefrac 1 4 - \nicefrac 2 3 \, \sW^2$, and for leptons $\gAl = -\nicefrac 1 4$ and $\gVl = -\nicefrac 1 4 + \sW^2$. Parity symmetry in the $t \bar t$ system is related to the values of $\gAt$ and $\gVt$:
\begin{align}
    \gAt = 0 \ \text{ or } \ \gVt = 0 \ &\implies \ \text{conservation of } P, \label{pcons}\\
    \gAt = \gVt \ \text{ or } \ \gAt = - \gVt \ &\implies \ \text{maximal violation of } P. \label{pmaxval}
\end{align}

\begin{figure}[H]
    \centering
    \includegraphics[width=.49\textwidth]{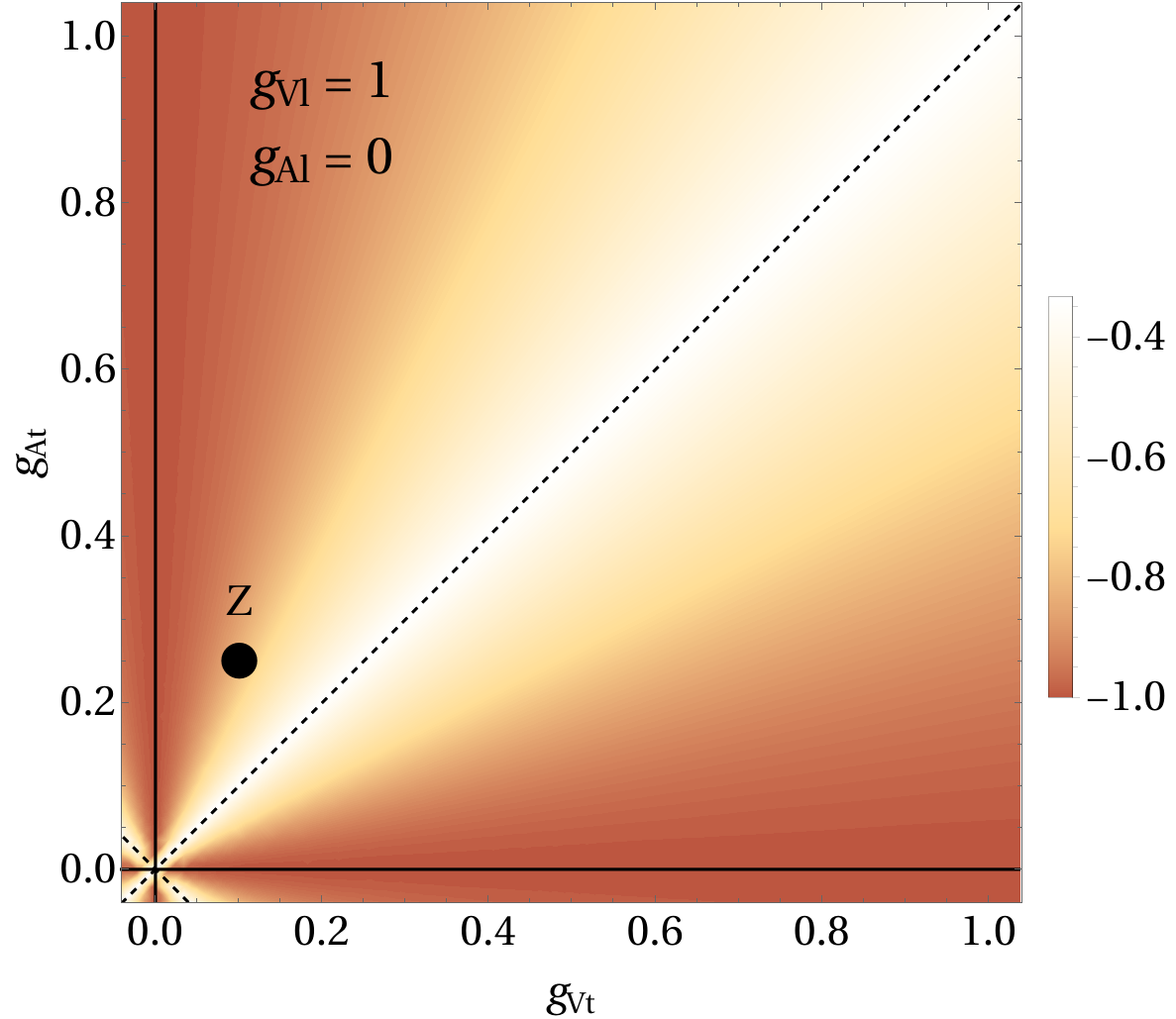}
    \includegraphics[width=.49\textwidth]{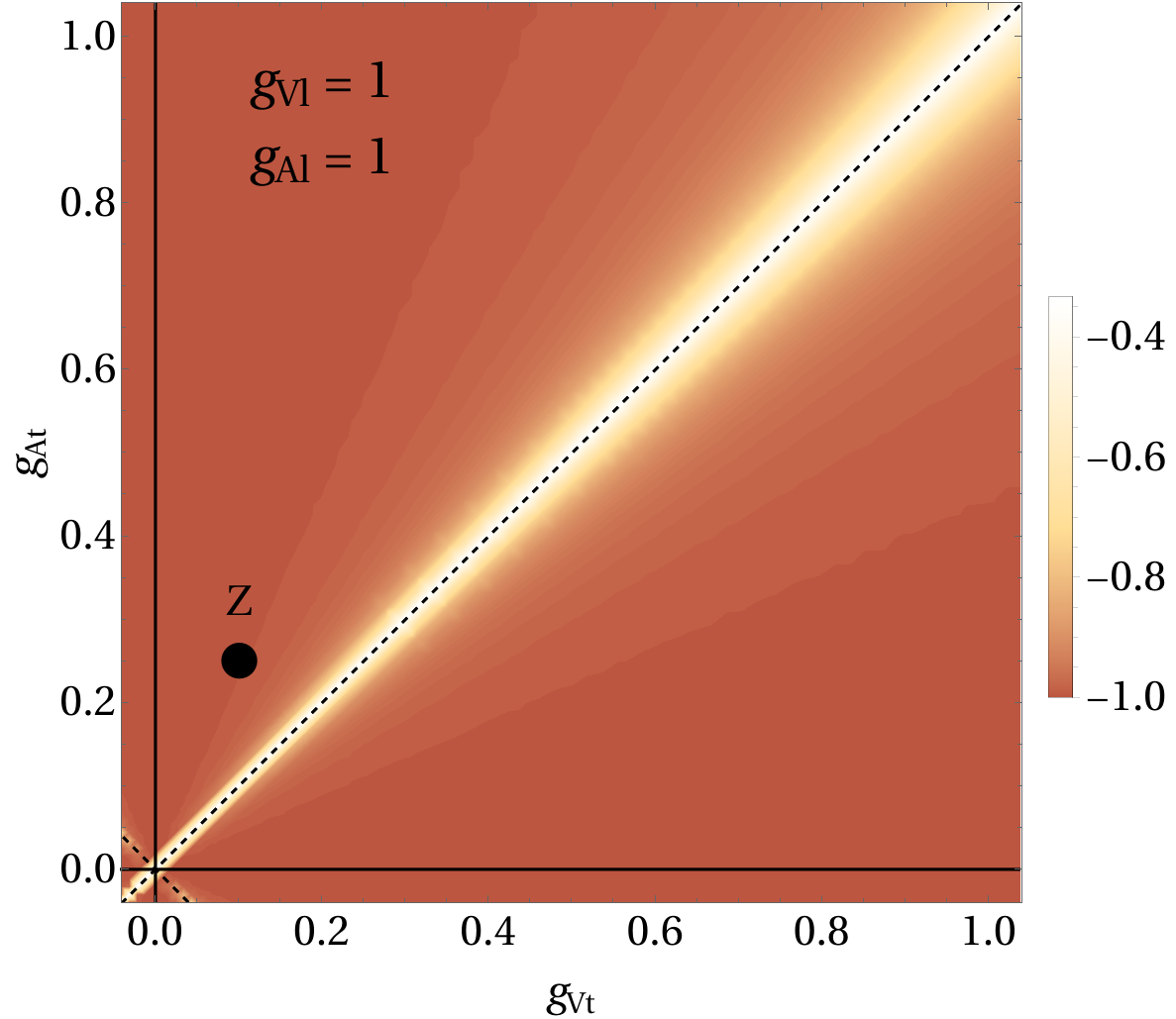} \\
    \vspace{5mm}
    \includegraphics[width=.49\textwidth]{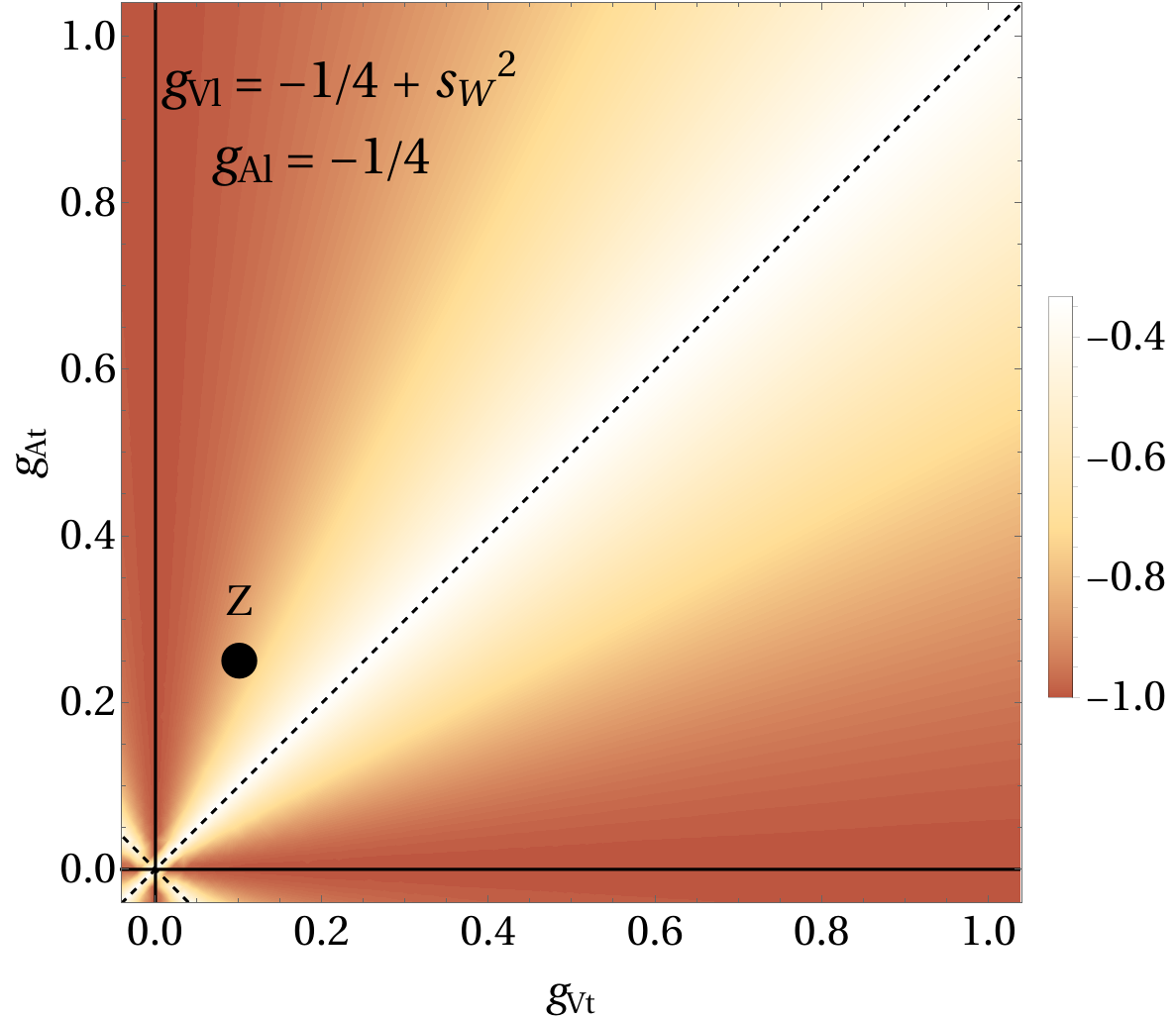}
    \includegraphics[width=.49\textwidth]{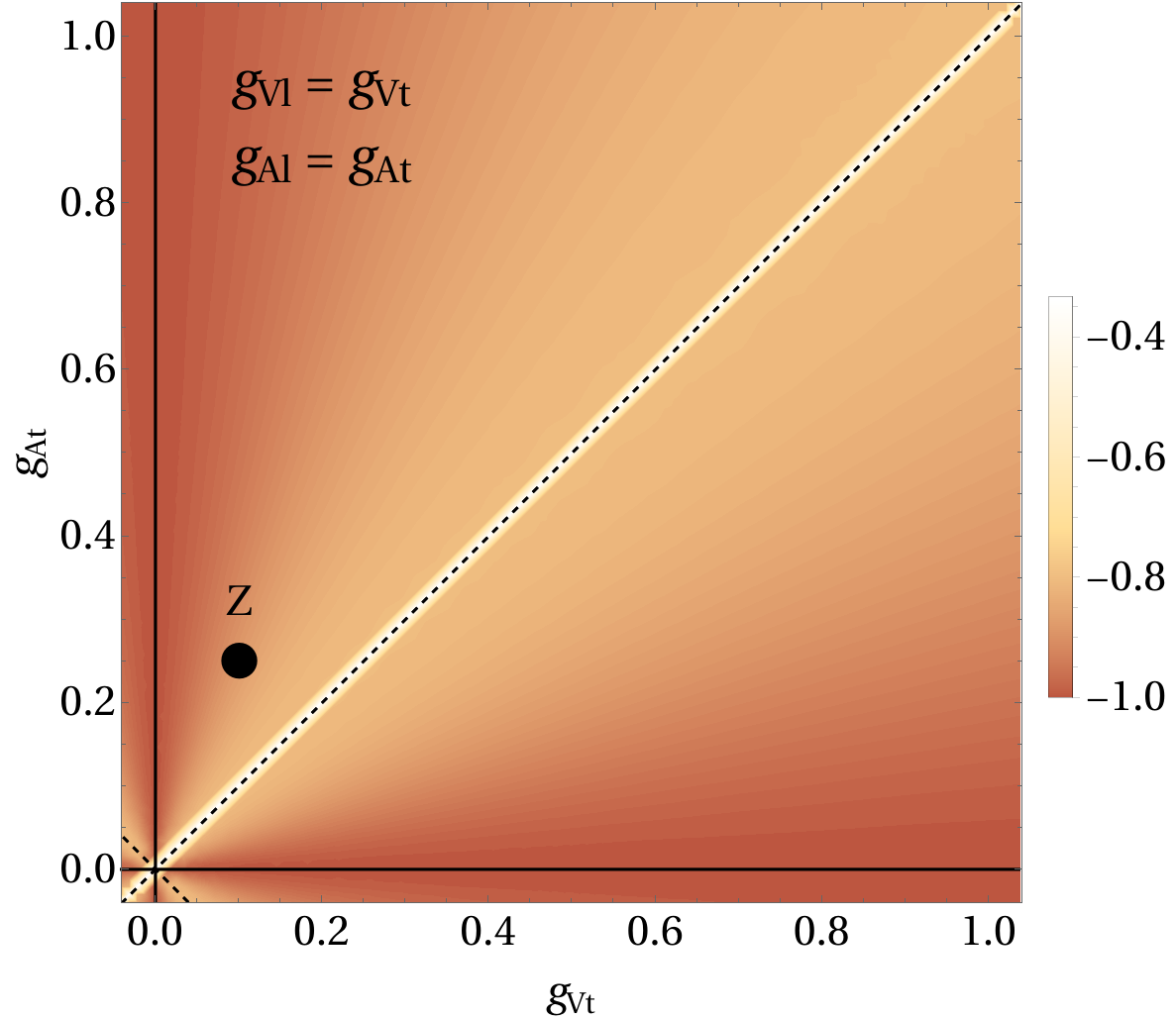}
    \caption{Value of $D_{\min}$ for different values of the vector and axial couplings $\gVt$ and $\gAt$ of the top quark, and of the initial state lepton $\gAl$ and $\gVl$. The actual SM couplings of the $Z$ boson to the top quark are indicated for reference. The marker $D$ is evaluated for $\beta = 1$ and for the scattering angle $\theta_{\max}$ that maximises entanglement. The interactions conserving $P$ \eqref{pcons} in the top couplings are shown by the solid lines, while those maximally $P$ violating \eqref{pmaxval} are shown by the dashed line. We find a correspondence between entanglement (more negative $D$) and parity symmetry. }
    \label{fig:gAt}
\end{figure}

For each value of $\gAt$ and $\gVt$, we then solve for the scattering angle $\theta_{\text{max}}$ that maximises the amount of entanglement. For instance, in the SM case shown in Figure \ref{fig:bell}, the optimal angle $\theta_{\text{max}}$ is approximately $0.7 \pi$. Having found the optimal configuration, we then evaluate the marker $D_{\min}$ of \eqref{dmin} that, as discussed above, is in one-to-one correspondence with entanglement of formation if off-diagonal terms of $\mathcal C$ are neglected. Entanglement is given for $D_{\min} < -\nicefrac 1 3$, Bell violations for $D_{\min} < -\nicefrac{1}{\sqrt 2}$, and maximal entanglement corresponds to $D_{\min} = -1$. In Figure \ref{fig:gAt} we show the value of $D_{\min}$ as a function of $\gAt$ and $\gVt$ for $\beta = 1$. \\

We find that, in all cases, the $P$--conserving configurations $k_\text{A} = 0$ and $k_\text{V} = 0$ reach a more negative, that is, more entangled, value of $D$ compared to $P$--violating configurations, showing a correspondence between entanglement and parity symmetry.

The general structure seen in Figure \ref{fig:gAt}, with maximal entanglement along the $P$--conserving configurations and reduced entanglement moving towards the maximally $P$--violating diagonal, is a general result, and is independent of the specific parameters in the problem, such as the mass of the top quark, the mass of the mediator boson(s), or the couplings between the vector boson and initial leptons.

\section{Spin correlations in the SMEFT} \label{sec:eft}

We now consider new physics contributions to $t \bar t$ production. Beyond-the-SM effects can produce a different spin density matrix $\rho$ compared to the SM prediction $\rho_{\text{SM}}$, while still being consistent with QM and QFT. Beyond-QM effects, on the other hand, have the opportunity of, say, yielding a $\rho$ with negative eigenvalues, or trace $\neq$ 1.  

Our study is within the framework of QFT, and by {\it new physics} we always mean new particles or interactions. If such new particles are heavy, their effect can be parameterized as a deformation of the SM in the form of higher dimensional operators, and we focus our attention to dimension-6 EFT operators that contain at least one top quark field, $Q_L$ and/or $t_R$. All operators enter the Lagrangian normalized by their Wilson coefficient and by a common cutoff parameter:
\begin{equation}
    \mathcal L \supset \frac{1}{\Lambda^2} \sum_i c_i \, \mathcal O_i.
\end{equation}

In the convention and notation of \cite{Aguilar-Saavedra:2018ksv, Degrande:2020evl}, there are three classes of dimension-6 operators entering $t \bar t$ production in a lepton collider at leading order:

\paragraph{Four fermion operators.}

These operators constitute the simplest class, as they directly mediate the process $\ell^+ \ell^- \to t \bar t$. In our convention they are given by:
\begin{align}
    \mathcal O_{Q\ell}^{(1)} &= (\overline Q_L \gamma^\mu Q_L) (\overline \ell_L \gamma_\mu \ell_L), \\
    \mathcal O_{Q \ell}^{(3)} &= (\overline Q_L \gamma^\mu \sigma_I Q_L) (\overline \ell_L \gamma_\mu \sigma^I \ell_L),\\
    \mathcal O_{Q e} &= (\overline Q_L \gamma^\mu Q_L) (\overline \ell_R \gamma_\mu \ell_R), \\
    \mathcal O_{t \ell} &= (\overline t_R \gamma^\mu t_R) (\overline \ell_L \gamma_\mu \ell_L), \\
     \mathcal O_{t e} &= (\overline t_R \gamma^\mu t_R) (\overline \ell_R \gamma_\mu \ell_R).
\end{align}
In all operators, the lepton field may be $\ell = e$ or $\mu$, depending on the lepton collider's nature. In this work we consider lepton colliders with unpolarized beams; for lepton collisions with significant initial-state polarization it may be useful introduce other SMEFT degrees of freedom, dividing the L and R lepton couplings \cite{Durieux:2018tev}.

To obtain simpler expressions, we will rotate the degrees of freedom of this sector as it is done commonly for $pp$ collisions, using as Wilson coefficients the set: 
\begin{align}
    &c_{Q \ell}^{(3)} + c_{Q \ell}^{(1)}, \\
    &c_{\VV} = \frac{1}{4} \big( c_{Q \ell}^{(1)} - c_{Q \ell}^{(3)} + c_{t e} + c_{t \ell} + c_{Q e}  \big), \\
    &c_{\AV} = \frac{1}{4} \big( -c_{Q \ell}^{(1)} + c_{Q \ell}^{(3)} + c_{t e} + c_{t \ell} - c_{Q e}  \big), \\
    &c_{\VA} = \frac{1}{4} \big( -c_{Q \ell}^{(1)} + c_{Q \ell}^{(3)} + c_{t e} - c_{t \ell} + c_{Q e}  \big), \\
    &c_{\AA} = \frac{1}{4} \big( c_{Q \ell}^{(1)} - c_{Q \ell}^{(3)} + c_{t e} - c_{t \ell} - c_{Q e}  \big). 
\end{align}

We note that, at LO, the only four-fermion degrees of freedom entering $t \bar t$ production are the usual $\VV$, $\VA$, $\AV$, $\AA$ \cite{Bernreuther:2015yna,Aoude:2022imd}, with $c_{Q \ell}^{(3)} + c_{Q \ell}^{(1)}$ only appearing at higher order.

\paragraph{Current current operators.} 

These operators contain two scalar and two fermion fields as two contracted currents, and modify the interaction of the $Z$ boson with the corresponding fermions. They are given by:

\begin{align}
    \mathcal O_{\phi Q}^{(1)} &= i (\phi^\dagger \stackrel{\leftrightarrow}{D}_\mu \phi) (\overline Q_L \gamma^\mu Q_L), \\
    \mathcal O_{\phi Q}^{(3)} &= i (\phi^\dagger \stackrel{\leftrightarrow}{D}_{\mu \, I} \phi) (\overline Q_L \gamma^\mu \sigma^I Q_L), \\
    \mathcal O_{\phi t} &= i (\phi^\dagger \stackrel{\leftrightarrow}{D}_\mu \phi) (\overline t_R \gamma^\mu t_R),
\end{align}
where $\ell$ may be $e$ or $\mu$. Similarly to the four-fermion sector, to highlight the chiral structure, we will use the Wilson coefficients:
\begin{align}
    &c_{\phi Q}^{(3)} + c_{\phi Q}^{(1)}, \\
    &c_\pV = \frac{1}{2} \big( c_{\phi t} + c_{\phi Q}^{(1)} - c_{\phi Q}^{(3)}\big), \\
    &c_\pA = \frac{1}{2} \big( c_{\phi t} - c_{\phi Q}^{(1)} + c_{\phi Q}^{(3)}\big).
\end{align}
Again, we note that $c_{\phi Q}^{(3)} + c_{\phi Q}^{(1)}$ does not enter $t \bar t$ production at LO. 

\paragraph{Dipole operators.} 

These operators give rise to an anomalous electroweak dipole moment of the top quark. They explicitly modify $t \bar t$ production at tree-level entering in the $ttZ$ or $tt\gamma$ vertex.
\begin{align}
    \mathcal O_{tW} &= (\overline Q_L \gamma^{\mu \nu} \sigma_I t_R) \, \widetilde \phi \, W^I_{\mu \nu}, \label{otw}\\
     \mathcal O_{tB} &= (\overline Q_L \gamma^{\mu \nu} t_R) \, \widetilde \phi \, B_{\mu \nu} \label{otb}.
\end{align}
Both $\mathcal O_{tW}$ and $\mathcal O_{tB}$ affect electroweak--mediated top pair production. In addition, the operator $\mathcal O_{tW}$ can change the properties of the top decay vertex $t W^+ b$. It is known, however, that the tops' spin state is transferred to their decay products even in the presence of non-zero electroweak dipoles \cite{Severi:2022qjy}. For the purposes of spin studies, therefore, it is enough to consider insertion of dipole operators in the $t \bar t$ production vertex. After spontaneous symmetry breaking, it is convenient to define the Wilson coefficients:
\begin{align}
    c_\tZ &= \cW \, c_{tW} - \sW \, c_{tB}, \\
    c_\tA &= \sW c_{tW} + \cW c_{tB},
\end{align}
so that the modified $t\bar t Z$ vertex is only a function of $c_\tZ$, and the modified $t \bar t \gamma$ vertex only depends on $c_\tA$.

\subsection{Analytical results at order $1/\Lambda^2$} \label{1L2}

We first consider the $\mathcal O(1/\Lambda^2)$ interference between SM and SMEFT amplitudes. Similarly to the SM case \eqref{a_SM}, we split the squared matrix element in two parts,
\begin{equation}
    A = A^{[6,0]} + A^{[6,1]},
\end{equation}
again counting the number of $\gamma^5$ insertion in the SM matrix element that is being interfered with the EFT amplitude. Explicitly, the $A^{[6,0]}$ contains terms proportional to $\Qt/\Lambda^2$ and $\gVt/\Lambda^2$, while $A^{[6,1]}$ is proportional to $\gAt/\Lambda^2$. \smallskip

In fact, for most SMEFT operators a calculation is not needed. Since the final $t \bar t$ spin state only depends upon the Dirac structure inserted in the top fermion line, for vector--like EFT operators, that is in the presence of:
\begin{equation}
    c_{\VV}, \ c_{\VA}, \ c_\pV, \label{cV}
\end{equation}
the quantum state reached in the $A^{[6,0]}$ channel is the same as the reached by the SM in the $A^{[0]}$ channel, in \eqref{EW_top_spin}, while the quantum state of the $A^{[6,1]}$ channel is the same as the SM $A^{[1]}$ channel, also in \eqref{EW_top_spin}. The only difference with the SM lies in the common factors, that here are given by \eqref{F60V} and \eqref{F61V}.  \smallskip

Similarly, for axial--like EFT operators, that is for:
\begin{equation}
    c_{\AV}, \ c_{\AA}, \ c_\pA, \label{cA}
\end{equation}
the quantum state reached in $A^{[6,0]}$ is identical to the SM $A^{[1]}$, while the state reached for $A^{[6,1]}$ is identical to the one obtained in the SM for $A^{[2]}$. The common factors $F$ are given by \eqref{F60A} and \eqref{F61A}.

The dipole operators \eqref{otw}--\eqref{otb}, and therefore the Wilson coefficients:
\begin{equation}
    c_\tZ, \ c_\tA, \label{cD}
\end{equation}
are the only terms in the SMEFT lagrangian capable of introducing a new Dirac structure in the top fermion line, and hence give rise to a spin state unreachable by any SM channel. It is given by:

\begin{equation}
\begin{dcases}
    A^{[6,0,\text{D}]} &=  F^{[6,0,\text{D}]}, \\
\widetilde{C}^{[6,0,\text{D}]}_{kk} &=  F^{[6,0,\text{D}]} \, c_\theta^2, \\
\widetilde{C}^{[6,0,\text{D}]}_{kr} &= F^{[6,0,\text{D}]} \frac{2 - \beta^2}{2\sqrt{1-\beta^2}} \, c_\theta s_\theta, \\
\widetilde{C}^{[6,0,\text{D}]}_{rr} &=  F^{[6,0,\text{D}]} \,s_\theta^2, \\
\widetilde{C}^{[6,0,\text{D}]}_{nn} &= 0,\,
\end{dcases}
\quad
\begin{dcases}
    A^{[6,1,\text{D}]} &= F^{[6,1,\text{D}]} \, c_\theta, \\
\widetilde{C}^{[6,1,\text{D}]}_{kk} &= F^{[6,1,\text{D}]} \, c_\theta, \\
\widetilde{C}^{[6,1,\text{D}]}_{kr} &= F^{[6,1,\text{D}]} \, \frac{1}{2 \sqrt{1-\beta^2}} s_\theta, \\
\widetilde{C}^{[6,1,\text{D}]}_{rr} &= 0, \\
\widetilde{C}^{[6,1,\text{D}]}_{nn} &= 0.\,
\end{dcases} \label{dipole_top_spin}
\end{equation}

The common factors for these channels are given by the expressions in \eqref{F60D} and \eqref{F61D}. The spin correlation matrices \eqref{dipole_top_spin} are singular for $\beta \to 1$, showing explicitly the EFT truncation. As we will see in the next Section, adding $\mathcal O(1/\Lambda^4)$ EFT-squared terms cures the divergence.

\subsection{Analytical results at order $1/\Lambda^4$}

Moving to the $\mathcal O(1/\Lambda^4)$ squared SMEFT amplitude, we find that, like the previous case, the spin states reached are for the most part already known.

For purely--vector interactions, that is for the square of an amplitude containing the coefficients \eqref{cV}, the state reached is the same as the $A^{[0]}$ channel in the SM, in \eqref{EW_top_spin}. The corresponding normalisation factor is in \eqref{F8VV}. Similarly, for purely--axial interactions, given by the square of \eqref{cA}, the resulting state is identical to the SM $A^{[2]}$, with the normalisation in \eqref{F8AA}. Mixed vector--axial interactions, given by the interference of an amplitude containing \eqref{cV} with another amplitude containing \eqref{cA}, give rise to the same spin state as the SM $A^{[1]}$, normalized by \eqref{F8VA}.

Purely--dipole interactions, given by the square of \eqref{cD}, are new at order $\mathcal O(1/\Lambda^4)$ and in fact give rise to a new spin state (very similar, but not identical to, the SM $A^{[0]}$):

\begin{equation}
\begin{dcases}
    A^{[8,\DD]} &= F^{[8,\DD]}\left( -\beta^2 c_{\theta }^2-\beta^2+2  \right), \\
\widetilde{C}^{[8,\DD]}_{kk} &= F^{[8,\DD]} \left[ -\beta^2+\left(2-\beta^2\right) c_{\theta }^2  \right], \\
\widetilde{C}^{[8,\DD]}_{kr} &= F^{[8,\DD]} \, 2 \sqrt{1-\beta^2} c_{\theta } s_{\theta } , \\
\widetilde{C}^{[8,\DD]}_{rr} &= F^{[8,\DD]} \, s_{\theta }^2 (2 -\beta^2), \\
\widetilde{C}^{[8,\DD]}_{nn} &=  F^{[8,\DD]}  \, \beta^2 s_{\theta }^2.
\end{dcases}
 \label{dipole2_top_spin}
\end{equation}

The corresponding factor $F^{[8,\DD]}$ is in \eqref{F8DD}. Mixed dipole--vector and mixed dipole--axial, given by the interference of an amplitude with \eqref{cD} with an amplitude with \eqref{cV} or \eqref{cA}, produce the same spin states we already found in \eqref{dipole_top_spin}, normalized by \eqref{F8VD} and \eqref{F8AD} respectively.

\subsection{Discussion} \label{EFTdiscussion}

As noted in the previous two Sections, the $t \bar t$ spin states reached by the SM and the SMEFT have a large degree of overlap. For clarity, we collect all the possible spin states reachable by the SM and the SMEFT in Table \ref{tab:spin}, noting the channels from which they are produced.

\begin{table}[h!]
\centering
\begin{tabular}{|c c |c |c| c |} 
\cline{3-5}
 \multicolumn{1}{c}{}& & \multicolumn{3}{c|}{$\mathcal M_1$} \\
 \multicolumn{1}{c}{}& & $\Qt$, $\gVt$, & $\gAt$, &  \\
 \multicolumn{1}{c}{}& &  $c_{\VV}$, $c_{\VA}$, $c_\pV$ & $c_{\AV}$, $c_{\AA}$, $c_\pA$ & $c_\tZ$, $c_\tA$ \\
     \hline
 \multirow{5}{*}{$\mathcal M_2$} & $\Qt$, $\gVt$
  & \multirow{2}{*}{$A^{[0]}$} & \multirow{2}{*}{$A^{[1]}$} & \multirow{2}{*}{$A^{[6,0,D]}$}\\ 
  &  $c_{\VV}$, $c_{\VA}$, $c_\pV$  & & &  \\
   \cline{2-5}
  & $\gAt$ & \multirow{2}{*}{$A^{[1]}$} & \multirow{2}{*}{$A^{[2]}$} & \multirow{2}{*}{$A^{[6,1,D]}$}\\ 
  & $c_{\AV}$, $c_{\AA}$, $c_\pA$ & & &  \\
  \cline{2-5}
  & & \multirow{2}{*}{$A^{[6,0,D]}$} & \multirow{2}{*}{$A^{[6,1,D]}$} & \multirow{2}{*}{$A^{[8,\DD]}$}\\ 
  &  $c_\tZ$, $c_\tA$ &  &  &  \\
  \hline
\end{tabular}
\caption{Spin state given by a process with matrix element squared $\mathcal M_1 \mathcal M_2^\star$, order by order in the couplings (SM or SMEFT) contained in the corresponding amplitudes.}
\label{tab:spin}
\end{table}

The simplicity of the results in Table \ref{tab:spin} allows us to draw some general conclusions, that are valid for $t \bar t$ production in a lepton collider both in the SM and in the dimension-$6$ SMEFT, at LO.

Firstly, the $D^{(1)}$ entanglement marker \eqref{D1} is fixed:
\begin{equation}
    D^{(1)} = \frac 1 3 \label{DfixSMEFT}
\end{equation}
(again note the $+$ sign).  This result, that we had already noticed for the SM in \eqref{DfixSM}, remains valid once dimension $6$ terms are included, and implies that the discovery power of $D^{(1)}$ for new physics is very limited.   

Next, the off diagonal symmetric correlations $C_{kn}+C_{nk}$ and $C_{rn}+C_{nr}$ vanish identically, 
\begin{equation}
    C_{kn}+C_{nk} = C_{rn}+C_{nr} = 0,
\end{equation}
again both in the SM and in the SMEFT. Overall, there are three degrees of freedom left in $t \bar t$ spin correlations: two of the three elements of the diagonal of $\mathcal C$, and the off-diagonal pair $C_{kr}+C_{rk}$. In the next Section, we will investigate their reach in searches for heavy new physics. \smallskip

We conclude this section with a remark on the high-energy limit in the EFT. In Section \ref{1L2} we had noted that the $\beta \to 1$ limit was pathological for $C_{kr}$ in the presence of dipole operators at order $1/\Lambda^2$. Recall that $C_{kr}$, being a physical correlation between two observables, is bounded by $1$, while in \eqref{dipole_top_spin} we see that the SMEFT prediction diverges. Of course, the general EFT assumption that $\Lambda$ is a heavy, unreachable scale, is violated when $s$ grows past $\Lambda^2$. Barring any difficulties in the interpretation of an EFT amplitude in the high-energy limit, here we simply note that the divergence in \eqref{dipole_top_spin} is mathematically cured by the inclusion of the $1/\Lambda^4$ terms we obtained in \eqref{dipole2_top_spin}. \\

\begin{figure}[H]
    \centering
    \includegraphics[width=.6\textwidth]{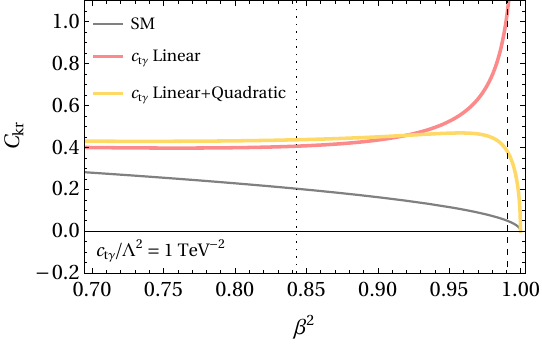}
    \caption{Value of $C_{kr}$ with the inclusion of the dipole operator $\mathcal O_{t\gamma}$, as a function of the top velocity $\beta$, and thus the center of mass energy. We indicated with a dotted line the point where the EFT contribution to $C_{kr}$ is as large as the SM background, and with a dashed line the critical point \eqref{critical}. The EFT prediction truncated at linear order $\mathcal O(c/\Lambda^2)$ diverges, while the curve at $\mathcal O(c/\Lambda^2 + c^2/\Lambda^4)$ remains physical for all $\beta$.}
    \label{fig:ckr}
\end{figure}

In Figure \ref{fig:ckr} we show the value of $C_{kr}$ for the SM and the SMEFT at linear and quadratic order, for:
\begin{equation}
    \frac{c_{t\gamma}}{\Lambda^2} = 1 \,\text{TeV}^{-2}.
\end{equation}
While at low energy ($\ll 1 \, \text{TeV}$) the quadratic contribution is indeed very small, we see that in the $\beta \to 1$ limit the prediction for $C_{kr}$ truncated at linear order diverges. The value of $C_{kr}$ becomes non-physical, that is, larger than $1$, around where:
\begin{equation}
    s = \Lambda^2, \quad \text{with} \quad c_{t \gamma}= 4 \pi. \label{critical}
\end{equation}
The full SMEFT prediction at order up to $1/\Lambda^4$, however, remains well-behaved regardless of energy.

\section{Searches for new physics using quantum observables} \label{sec:analysis}

We have shown recently \cite{Maltoni:2024tul} that at the LHC in numerous notable NP scenarios, quantum observables such as the entanglement markers \eqref{D1}-\eqref{D4} provide complementary information to classical observables, and their inclusion in a search may yield the additional sensitivity needed to detect new phenomena.

In a similar way, in this Section we investigate the sensitivity of $t \bar t$ spin observables to non--resonant heavy new physics in future lepton colliders. We first consider in detail a scenario that is commonly considered in the literature, and then we show more general results for an energy scan from $t \bar t$ threshold up to the TeV scale.

The specific scenario we consider is a $365 \, \text{GeV}$ circular electron--positron collider, collecting $1.5 \, /\text{ab}$ of luminosity \cite{FCC:2018byv,FCC:2018evy,FCC:2018vvp,Benedikt2020}. Because of the clean environment given by a lepton collider, it is realistic that both the dilepton and the lepton+jet decay channels will be available for top spin studies, and we therefore include both in our simulations. Our estimates for the number of events usable for physics is collected in Table \ref{tab:Ntop}.

\begin{table}[h!]
\centering
\begin{tabular}{|c |c|}
\hline
 & $e^+ \, e^-$  365 GeV \\
 \hline
 Luminosity & $1.5 \, \text{/ab}$ \\
 Partonic cross-section & $750 \, \text{fb}$ \\
    \hline
 Events at parton level & $1\,125\,000$ \\
 Reconstructed events & $450\,000$ \\
\hspace{8mm} \rotatebox[origin=c]{180}{$\Lsh$} lepton+jet channel & $135 \, 000$  \\
 \hspace{4mm} \rotatebox[origin=c]{180}{$\Lsh$} dilepton channel & $22 \, 500$ \\
 \hline
\end{tabular}
\caption{Estimated yield of $t \bar t$ events for the lepton collider scenario we are considering. Reconstruction of an event assumes a global reconstruction efficiency of $40 \%$ \cite{ATLAS:2023fsd}, and final states with a $\tau$ lepton of either sign are not included.}
\label{tab:Ntop}
\end{table}

We simulate an inclusive measurement of the top/anti-top spin correlation matrix $\mathcal C$ and of its entanglement markers $D$. The observables we consider are:
\begin{enumerate}[itemsep=0pt]
    \item The number of events $N$.
    \item The entanglement markers $D^{(k)}$ and $D^{(n)}$, that parameterize the diagonal of $\mathcal C$.
    \item The average off-diagonal spin correlation $C_{kr}+C_{rk}$.
\end{enumerate}

Our choice of observables is based on two considerations: first, we expect that in a lepton collider the entries of $\mathcal C$ will be measurable much more easily than in a $pp$ machine, therefore we directly base our searches for new physics on the reconstructed $C_{ij}$, rather than on proxy observables such as $\Delta \phi_{\ell \ell}$. Further, from the general results we obtained in Section \ref{EFTdiscussion}, we removed all observables that are not expected to yield any SMEFT sensitivity at LO: $D^{(1)}$, $D^{(r)}$ (that is fixed by \eqref{DfixSMEFT} and by knowledge of $D^{(k)}$ and $D^{(n)}$) and the $kn$ and $rn$ off-diagonal terms of $\mathcal C$.

The inclusion of spin measurements such as those we consider in this study will prove particularly powerful when combined with the plethora of studies based on kinematics observables, such as $p_T$ and $m_{t \bar t}$ distributions, as they probe different and complementary degrees of freedom. In the absence of systematic uncertainty, the problem of devising quantities that extract the maximal amount of information from a sample is solved, and results in the so-called {\it optimal observables} \cite{Diehl:1993br}. Optimal observables that have been presented in the literature \cite{DeBlas:2019qco,deBlas:2022ofj,Celada:2024mcf} can readily be extended to include the spin of top quarks by considering tops' decay products, as electroweak decays turn a measurement of spin into a measurement of angles. However, since all measurements of spin correlation and entanglement with top quarks are currently systematics-dominated, and estimating systematical uncertainties in a future lepton collider is far from trivial, we elected to only employ observables whose usability has already been proven at the LHC by ATLAS and CMS.

\subsection{Estimation of uncertainties}

We assign to each measurement $x$ a Poisson statistical
uncertainty $\pm \Delta x$ of the form:
\begin{equation}
    \Delta x = x \, \frac{\kappa}{\sqrt{N}}, \label{stat}
\end{equation}
where $\kappa$ is a parameter of order one, and $N$ is the recorded number of events at detector level.

In the dilepton channel, the $t \bar t$ state is reconstructed via the angular distributions of charged leptons, that have maximal spin analyzing power,
\begin{equation}
    \alpha_\ell = 1.
\end{equation}
In the lepton+jet channel, the spin of one top quark is reconstructed from the charged lepton, while the spin of the hadronically-decaying top quark is reconstructed from the two light QCD jets. In the absence of light--flavor tagging, there exists an optimal combination \cite{Tweedie:2014yda} of the two light jet momenta that yields a spin analyzing power of approximately:
\begin{equation}
    \alpha_j = 0.64. \label{alphaj}
\end{equation}
One therefore expects a mild worsening in the $\kappa$ factors of \eqref{stat} in the lepton+jet channel because $\alpha_j < \alpha_\ell$. Conversely, however, the relative branching ratios $\text{BR} (t \bar t \to \ell \ell)$ and $\text{BR}(t \bar t \to \ell j)$ improve by a factor of approximately $6$. Overall, we expect the significance for a discovery to approximately scale as:
\begin{equation}
    ( \text{significance lepton+jet} ) \approx (\text{significance dilepton} ) \times \frac{\alpha_j}{\alpha_\ell} \sqrt{\frac{\text{BR} (t \bar t \to \ell j)}{\text{BR}(t \bar t \to \ell \ell)}}.
\end{equation}

As it is commonly done in lepton collider studies, we will also assume that systematic uncertainties are negligible with respect to statistical ones, especially considering the cleanness of $t \bar t$ events. Table \ref{tab:unc} collects our final numerical estimates for the uncertainties in the dilepton and lepton+jet channels. \smallskip

\begin{table}[ht]
\centering
\begin{tabular}{|c c c|} 
 \hline
 Observable & Uncertainty ($\ell \ell$) & Uncertainty($\ell j$)\\
 \hline\hline
 $N$ & $\sqrt{N_{\ell \ell}}$ & $\sqrt{N_{\ell j}}$ \\ 
  \hline
$D^{(k,n)}$ & $0.75 / \sqrt{N_{\ell \ell}}$ & $0.75 / (\alpha_j \sqrt{N_{\ell j}})$ \\
 \hline
 $C_{kr}+C_{rk}$ & $3.0 / \sqrt{N_{\ell \ell}}$ & $3.0 / (\alpha_j \sqrt{N_{\ell j}})$ \\
 \hline
\end{tabular}
\caption{Estimated absolute statistical uncertainty on the observables considered in this work as a function of the number of events collected, for the dilepton and lepton+jet final states, adapted from similar measurements at the LHC \cite{ATLAS:2023fsd,CMS:2019nrx}. }
\label{tab:unc}
\end{table}

\subsection{Monte Carlo generation details}

In the previous Sections we presented the short-distance structure of spin correlations in \eqref{EW_top_spin}. We note, however, that phenomenological predictions for colliders require additional ingredients.  Firstly, the evolution of the lepton beams has to be modelled appropriately. Unlike proton PDFs, that have an intrinsic non-perturbative nature and must therefore be extracted from data, the evolution of leptons can be evaluated analytically, and several public libraries exist for this purpose \cite{Frixione:2019lga,Bertone:2019hks,Frixione:2012wtz,Bertone:2022ktl,Frixione:1993yw,Ruiz:2021tdt,Shao:2022cly}. Remaining below the TeV scale, however, we checked that the inclusion of lepton PDFs does not yield significantly different results compared to assuming no beam evolution at all. Secondly, parton-shower effects of QCD as well as EW origin should be taken into account. We remark that the modelling of soft QCD radiation has proven to be very challenging in top spin measurements at the LHC \cite{ATLAS:2023fsd, CMS:2024pts}. However, we do not expect an equally large contribution (and therefore modelling uncertainty) to appear in a lepton collider, and it is reasonable that, thanks to accurate parton-shower simulations, particle-level measurements will be successfully unfolded to parton-level.

In the following, we will present simulated analyses unfolded to parton--level, based on {\tt MadGraph5\_aMC@NLO} \cite{1405.0301}, that has recently been equipped with support for lepton-lepton collisions \cite{Frixione:2021zdp}. Electron--positron collisions are simulated at LO accuracy in the electroweak couplings. The renormalisation and factorisation scales are set to $\mu_R = \mu_F = \sqrt{s}/2$. Our simulation of the SM background contains all tree-level diagrams at order $\alpha^6$ for top pair production and subsequent decay into the desired final state, composed of two $b$ jets, as well as two charged leptons for the dilepton channel, and one charged lepton and one light jet for the lepton+jet channel. In all cases we remove events containing at least one $\tau$ lepton of either sign. The NP signal is generated from the {\tt SMEFT@NLO} UFO model \cite{Degrande:2020evl}, including the operators we considered in Section \ref{sec:eft}. The MC simulation includes the interference between SM and EFT amplitude at order $1/\Lambda^2$, as well as the square of the EFT dimension--6 amplitude at order $1/\Lambda^4$.

\subsection{Results for $365 \, \text{GeV}$} \label{sec:365}

We extract constraints with a $\chi^2$ test using the observables described above, considering two Wilson coefficients at a time. In the following Figures \ref{fig:fit_1}-\ref{fig:fit_n} we showcase our results with a few representative plots. All the remaining plots are available as supplementary material to the online version of this manuscript.

\begin{figure}[ht]
    \centering
    \includegraphics[width=.85\textwidth]{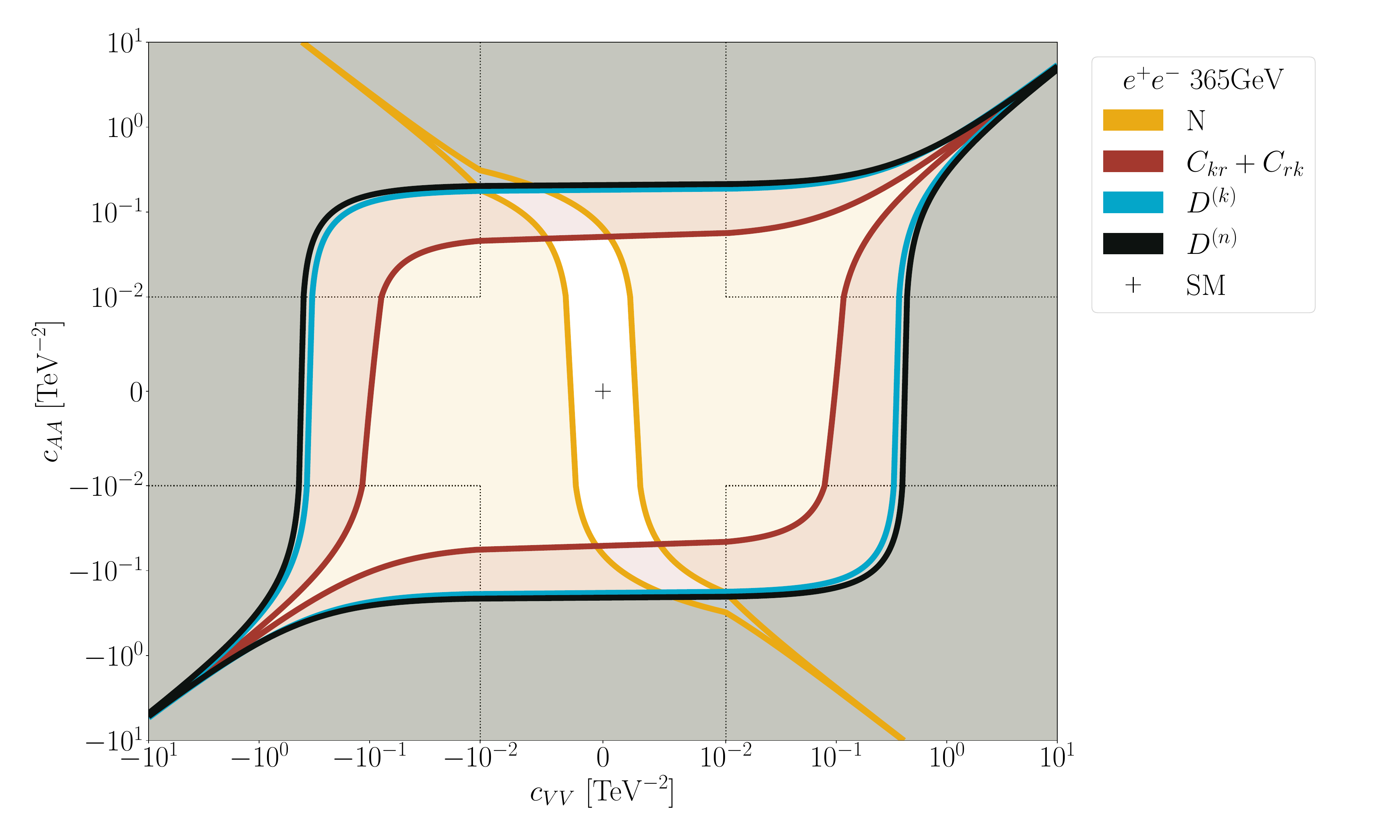} \\
    \caption{Regions in the $c_\VV-c_\AA$ plane expected to be excluded at $2 \sigma$ from each one of the observables we considered in our study, from the combination of $t \bar t$ dilepton and $t \bar t$ lepton+jet channels. The excluded regions are shaded, and outside the colored lines. The fit shown is at linear order in $c/\Lambda^2$, quadratic effects are negligible. The scale is logarithmic outside the dashed lines, and linear in the region that crosses zero.}
    \label{fig:fit_1}
\end{figure}

\begin{figure}[t]
    \centering
    \includegraphics[width=.85\textwidth]{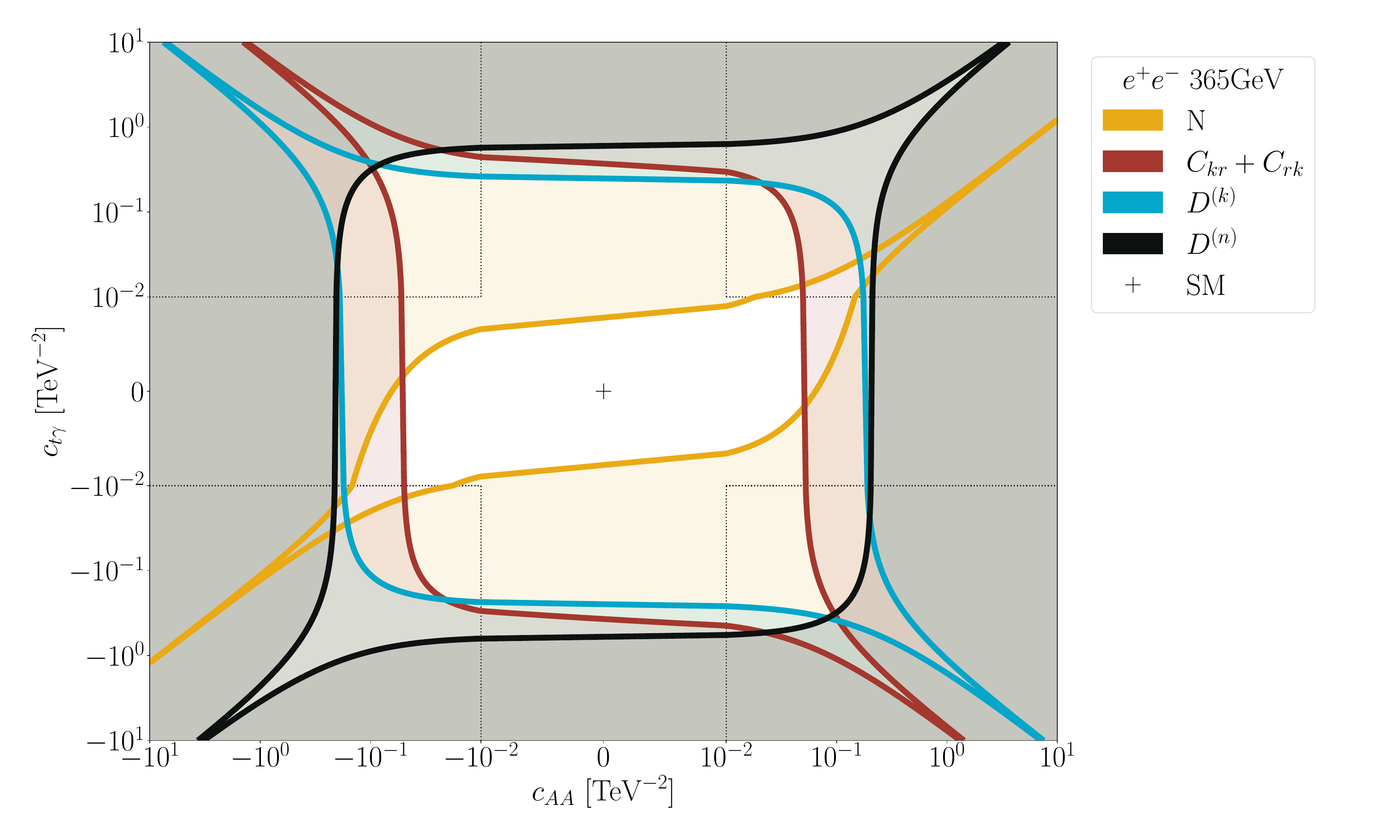} \\
    \caption{Same as Figure \ref{fig:fit_1} but for $c_\AA$ and $c_\tA$.}
    \label{fig:fit_n}
\end{figure}

We first note that quadratic effects in the fit are very small, and the constraints we obtain at order $1/\Lambda^2$ (that are shown in the figures) are basically identical to those at order $1/\Lambda^4$, obtained when including the squared dimension-6 contributions.

As expected, the most promising observable in constraining SMEFT Wilson coefficients is the number of events, corresponding to the total cross-section. However, it is a general fact that the simultaneous extraction of $n$ parameters from one observable will result in at least $n-1$ flat directions, along which the parameters only enter at higher order. The combination of different observables, especially when they explore different physical properties of the system under study, allows for a simple and theoretically clean way to address flat directions. We address this issue in detail in the next Section.

\subsection{Lifting of flat directions}

In our simulated analysis we find that, frequently, spin observables carry complementary information with respect to kinematics, and the flat directions corresponding to the total cross-section are orthogonal to the flat directions corresponding to spin. The combination of classical observables, based on $t \bar t$ kinematics, and spin observables, then reduces the number of flat directions, and it may even yield no flat directions at all.

This can be better understood by considering a pair of Wilson coefficients $(c_1, c_2)$ and evaluating explicitly the flat direction:
\begin{equation}
    c_1 = (\tan \alpha) \, c_2,
\end{equation}
for a variety of observables. Starting from the total rate $\sigma$, the SMEFT shift at linear order in $1/\Lambda$ is given by:
\begin{equation}
    \Delta \sigma = c_1 \, A^{[1]} + c_2 \, A^{[2]}.
\end{equation}
Clearly, then,
\begin{equation}\label{eq:flatangleN}
    \tan \alpha_{N} = -\frac{A^{[1]}}{A^{[2]}},
\end{equation}
is a flat direction when using $\sigma$ (that is, the number of events $N$) as observable. Let's now consider a quantum observable, call it $D$, defined as an angular average, {\it i.e.\@} a ratio of cross sections. In this case, schematically, the SMEFT shift is given by:
\begin{equation}
\Delta D_{\rm SMEFT} = \frac{\widetilde{D}^{[\rm SM]} + c_1 \widetilde{D}^{[1]} + c_2 \widetilde{D}^{[2]}}{A^{[\rm SM]} + c_1 A^{[1]} + c_2 A^{[2]}} - \frac{\widetilde{D}^{[\rm SM]}}{A^{[\rm SM]}}.
\end{equation}
A calculation shows that $\tan \alpha$ in this case is given by:
\begin{align}
\tan \alpha_D &= \frac{ \widetilde{D}^{[1]} A^{[\rm SM]} - A^{[1]} \widetilde{D}^{[\rm SM]}}{A^{[2]} \widetilde{D}^{[\rm SM]} - \widetilde{D}^{[2]} A^{[\rm SM]}}, \\
    & = \tan \alpha_{N} \cdot \frac{ D^{[1]} - D^{[\rm SM]} }{  D^{[2]}-D^{[\rm SM]}  }.
\label{eq:flatangleQ}
\end{align}
The flat direction $\alpha_D$ obtained using $D$ is in general different from the flat direction obtained with $N$. The best-case scenario in terms of discovery power happens when the flat directions \eqref{eq:flatangleN} and \eqref{eq:flatangleQ} are perpendicular, which happens for:
\begin{equation}
        \frac{ D^{[1]} - D^{[\rm SM]} }{  D^{[2]}-D^{[\rm SM]}}=-\left(\frac{A^{[2]}}{A^{[1]}}\right)^2.
\end{equation}
On the other hand, if $D^{[1]}=D^{[2]}$ the angles $\alpha_D$ and $\alpha_N$ are the same, and spin observables do not improve the SMEFT coefficient extraction.

In several cases, the off-diagonal spin correlation $C_{kr}+C_{kr}$ (red contour in the Figures) alone is enough to lift the degeneracy in the 2D plane stemming from the total rate $N$ (yellow). In other cases, like for the simultaneous fit of $(c_\VV,c_\tA)$, all spin observables are needed to effectively address the degeneracy. We also find a limited number of cases, like the combination $(c_\VA,c_\pV)$, where the degeneracy stemming from the total rate is not lifted by spin, and different approaches will be needed.  We remark that in the vast majority of cases spin observables are enough to lift flat directions.

\subsection{Collider energy scan} \label{sec:10000}

Proposed future colliders, with energies ranging from the $t\bar t$ threshold, investigated in Section \ref{sec:365}, to a futuristic $10+$ TeV muon accelerators, will be crucial in exploring the precision and energy frontier in this century. Such machines will also give indirect access to states significantly heavier than the corresponding collision energy, by exploiting the different scaling of the SMEFT and SM terms. At dimension six, the SMEFT interference cross section $\sigma_{\text{SMEFT}}$ can remain constant with the center of mass energy, $\sigma_{\text{SMEFT}} \sim 1/\Lambda^2$, decrease like $\mt / (\Lambda^2 \, \sqrt{s})$, or like $\mt^2 / (\Lambda^2 \, s)$. The SM cross-section, the background in this case, always decreases as $1/s$. This simple argument shows that scales of order $\Lambda$ can be explored with a discovery power that parametrically scales like:
\begin{equation}
    \frac{S}{\sqrt{B}} \sim \begin{dcases}
        \sqrt{s}/\Lambda^2, &\text{if } \sigma_{\text{SMEFT}} \sim 1/\Lambda^2, \\
        \mt/\Lambda^2 &\text{if } \sigma_{\text{SMEFT}} \sim \mt / (\Lambda^2 \, \sqrt{s}), \\
        \mt^2/(\Lambda^2 \, \sqrt{s}) &\text{if } \sigma_{\text{SMEFT}} \sim \mt^2 / (\Lambda^2 s).
    \end{dcases} \label{soverrootb}
\end{equation}

In the best-case scenario, it is not unrealistic for a $10 \, \text{TeV}$ muon collider to eventually explore scales $\Lambda$ well into the $100 \, \text{TeV}$, an unprecedented reach for models with a reasonable flavor structure. Note that in the worst-case scenario where SM and SMEFT decrease equally rapidly,  the discovery power actually worsens at increasing collider energy, due to the overall reduction in the number of events to extract the signal from.

The scaling considerations we just made apply to general-purpose kinematical observables, proportional to the total cross-section for the process. The case for quantum observables is more complicated, as such quantities depend on a combination of cross-sections, and therefore in general grow differently than our simple power counting in \eqref{soverrootb}. In practice this is not a problem, since, as we have shown, that the value of quantum observables is not in their specific exclusion bounds, but in the lifting of flat directions left over by other analyses. \smallskip

To appreciate the SMEFT discovery power scaling \eqref{soverrootb} and the corresponding breaking of flat directions induced by the inclusion of spin measurements, we extended the simulated analysis of Section \ref{sec:365} increasing the $\ell^+ \ell^-$ center of mass energy from $365 \, \text{GeV}$ up to the multi-$\text{TeV}$, and extracted:
\begin{itemize}
    \item The bound for Wilson coefficients obtained by only using the total rate $N$.
    \item The directions in SMEFT parameter space explorable by the spin observables $D^{(k)}$, $D^{(n)}$, and $C_{kr} + C_{rk}$, to compare with the flat direction left by $N$.
\end{itemize}

In the remainder of this Section, we show a selection of representative 2-dimensional SMEFT fits. The full set of plots is available as supplementary material to the online version of this work.

\begin{figure}[ht]
    \centering
    \includegraphics[width=.8\textwidth]{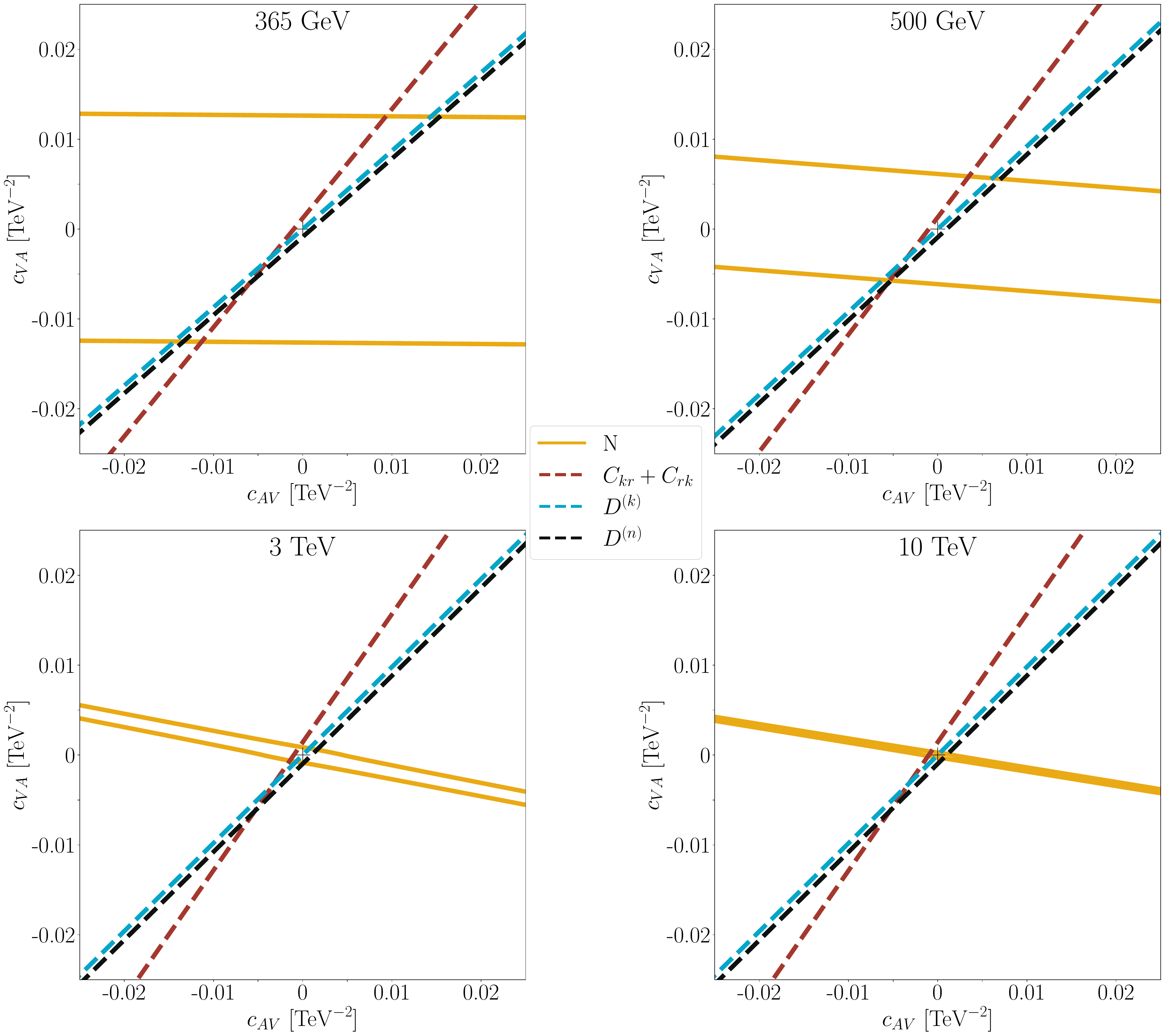} \\
    \caption{Exclusion region ($95 \%$ CL) in the $(c_\AV,c_\VA)$ plane, and flat directions corresponding to the spin observables $D^{(k,n)}$ and $C_{kr}+C_{rk}$, for a selection of collider energies: $365 \, \text{GeV}$, $500 \, \text{GeV}$, $3 \, \text{TeV}$, and $10 \, \text{TeV}$.}
    \label{fig:Energy-ss}
\end{figure}

The pair of Wilson coefficients of Figure \ref{fig:Energy-ss} exhibits a maximal energy growth, so that the discovery power scales approximately as $\sqrt{s}$, and indeed we find that the width of the yellow exclusion region decreases linearly with the center of mass energy. A flat direction is clearly seen, at an angle of about $ - 14^\circ$ with the $c_\AV$ axis. Spin observables also exhibit flat directions (in this case, at angles between $\approx 55 ^\circ$ and $45^\circ$). The large difference in angles means that combining kinematical and quantum observables will likely be very effective in lifting the individual flat directions.

\begin{figure}[ht]
    \centering
    \includegraphics[width=.8\textwidth]{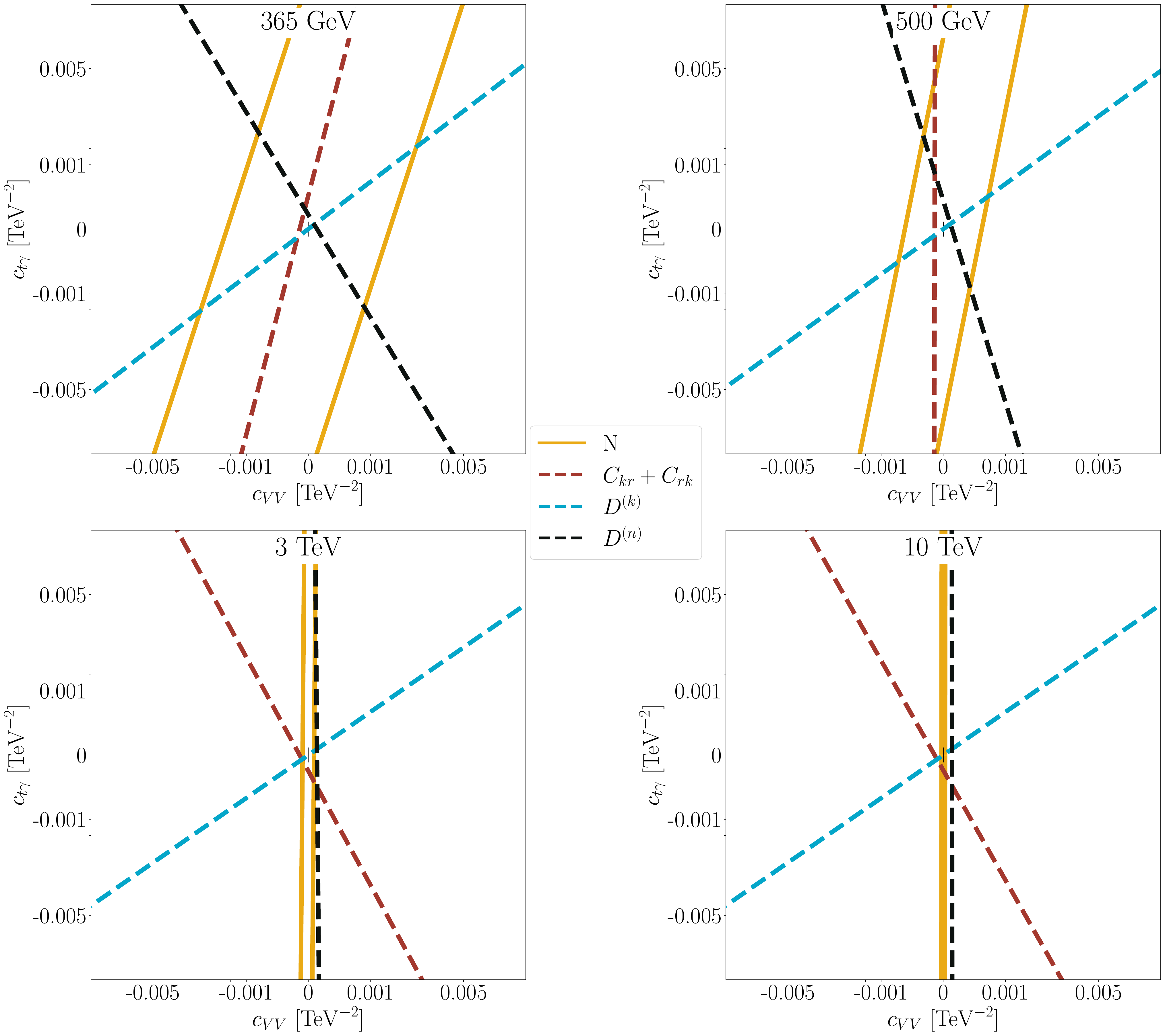} \\
    \caption{Same as Figure \ref{fig:Energy-ss} but for $c_\AV$ and $c_\tZ$.}
    \label{fig:Energy-snos}
\end{figure}

\begin{figure}[ht]
    \centering
    \includegraphics[width=.8\textwidth]{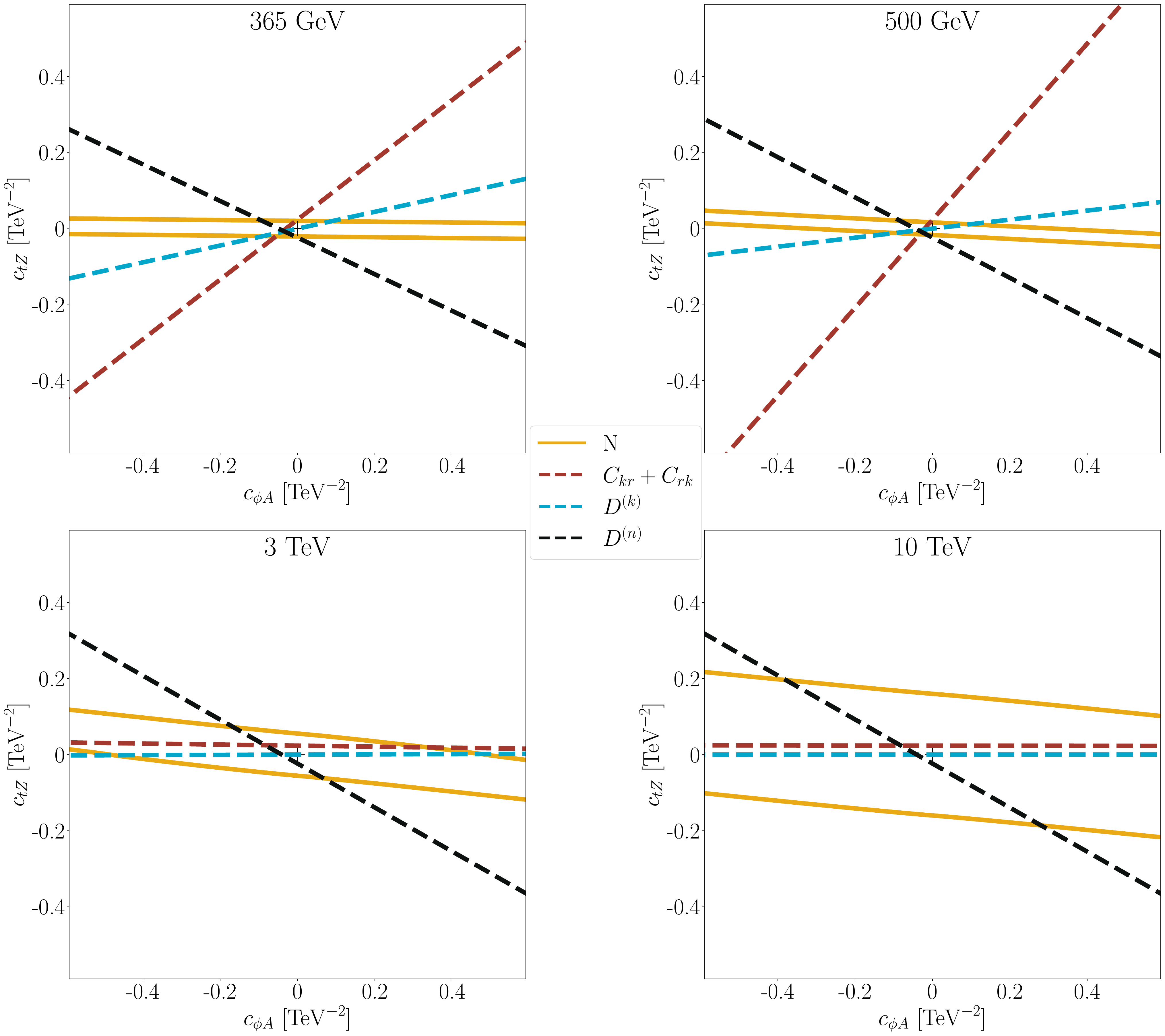} \\
    \caption{Same as Figure \ref{fig:Energy-ss} but for $c_\tZ$ and $c_\pA$.}
    \label{fig:Energy-nosnos}
\end{figure}

The case of two maximally growing amplitudes, like $c_\AV$ and $c_\VA$ is not the only one to consider. In Fig.~\ref{fig:Energy-snos} we consider a pair of Wilson coefficients whose corresponding contributions exhibit a different energy growth, constant for $c_\VV$ and decreasing as $1/s$ for $c_\tA$. The corresponding exclusion regions in Fig.~\ref{fig:Energy-nosnos} reflect such difference. The allowed range for $c_\VV$ (at constant $c_\tA$) improves linearly with energy, while the allowed range for $c_\tA$ (at constant $c_\VV$) is also linear in energy, but becomes worse and worse.

In any case, a flat direction clearly emerges when  considering $N$ as the only observable, and, as above, the inclusion of spin observables solves the problem. It is interesting to note that the flat direction for $D^{(n)}$ is almost orthogonal to the flat direction of $N$ at low energy, but asymptotically becomes aligned to it. The flat directions of $D^{(k)}$ and $C_{kr}+C_{rk}$ do not exhibit such behaviour, and remain well separated at all scales.

The last representative case we consider here is when both SMEFT contributions behave in the same way as the SM background. As we noted above this is a worst case scenario, that only takes place when all $1/\Lambda^2$ factors in the SMEFT amplitudes are saturated with positive powers of SM scales ({\it e.g.\@} the Higgs vev $v$ or the top mass $\mt$). The power-counting expectation for this case is for the discovery power to globally worsen with energy, and indeed this is what we find, see Fig.~\ref{fig:Energy-nosnos}. It is also interesting to note that in this particular case, there is not a clear improvement in flat directions at large energy, with only $D^{(n)}$ able to approximately lift the horizontal flat direction, while $N$, $C_{kr}+C_{rk}$, and $D^{(k)}$ all become degenerate.

\clearpage

\section{Conclusions} \label{sec:conclusion}

Lepton collisions at  the electroweak scale and above will provide an unprecedented insight into the properties of the Standard Model. One option under consideration is the construction of a $\sim 100$ kilometer $e^+ e^-$ accelerator (FCC-$ee$ or CEPC), that would be capable of reaching the energy required for direct $t \bar t$ production. It is anticipated that such a collider could record approximately one million $t \bar t$ events during its operation, and significantly refine our understanding of the top sector of the SM, yielding a significantly better precision on almost all measurements at the electroweak scale, compared to the HL-LHC (regardless of the eventual final luminosity).

In this work we have explored the phenomenology of a class of observables that relies on the quantum correlations between the spins of the top and anti-top quarks, in a $365 \, \text{GeV}$ lepton collider and in other future circular lepton-lepton machines. In our opinion these ``quantum'' measurements carry a double interest: as tests of fundamental aspects of QM at short distance, and as novel ways to search for new physics, {\it i.e.}\@ new particles and interactions.
We first reviewed spin correlations and entanglement markers for the top--anti-top qubit system, and then focused on top pair production in lepton collisions within the SM. We provided analytic expressions for the spin correlation coefficients and discussed how in unpolarised lepton lepton collisions the $t\bar{t}$ quantum spin state has an approximate optimal axis where spin correlations are maximal.  

We then studied in detail the quantum properties of top quark pairs in the SM, including entanglement and Bell violations, especially in conjunction with the specific properties of the electroweak interactions, that, unlike QCD,  break parity symmetry. We computed the spin density coefficients and entanglement markers at different collider energies, finding that the entanglement presence criterion is always satisfied for colliders running above the $t\bar{t}$ threshold. For Bell violation we showed that prospects for observing it improve at larger centre-of-mass energies and in specific phase space regions but would still require percent level experimental precision. Finally, we demonstrated how maximal entanglement aligns with the direction of parity conservation, whilst for maximal $P$-violation entanglement is reduced.  

Due to its importance for lepton colliders running close to the $t\bar{t}$ threshold we also explored the physics associated to the formation of top--anti-top bound states and their impact on spin correlation observables. We showed that in lepton collisions threshold effects as computed in non-relativistic QCD can be mimicked by introducing a vector resonance with an appropriate mass and width. As at the LHC, bound state effects modify the $t\bar{t}$ lineshape by increasing the total rate close to the threshold. However, at variance with $pp$ collisions, we confirmed the expectation that they have no effect on the entanglement markers as the vector resonance has identical spin quantum numbers as the photon and $Z$ mediating the process. 

Next, we studied how the SM picture changes under the inclusion of heavy new physics, for our purposes parameterised by dimension-6 SMEFT operators. We computed the spin density matrix analytically both at linear and quadratic order in the Wilson coefficients of relevant two-fermion and 4-fermion operators entering top pair production. Through a simple statistical analysis we explored the prospects of using quantum observables to constrain these Wilson coefficients at the FCC-ee. Similarly to the LHC case \cite{Severi:2022qjy}, we find that entanglement-inspired quantities provide complementary information to classical observables, which in turn helps alleviate some of the inherent degeneracies typical of the SMEFT. Specifically, by conducting simplified two-parameter fits, we explicitly confirmed that quantum observables lift flat directions that would otherwise appear when only considering the $t \bar t$ kinematics.

Additionally, for the sake of a more comprehensive analysis, and most importantly as a reference for future investigations, we considered higher collider energies, well beyond the production threshold and into the region where top quarks may be approximated as massless fermions. In such cases, the production cross-section and consequently the event count are expected to decrease as $1/s$. However, the sensitivity to new physics may increase, especially for the dimension-six operators whose effects amplify with energy. We thoroughly evaluated these effects and determined how the EFT space directions probed by the spin observables are modified at different collider energies. 

We view our study as an initial exploration, which can be refined and extended in several directions. Firstly, it is important to note that our analysis was conducted at the tree level, without accounting for higher-order corrections on quantum correlations arising from either strong or electroweak interactions. Secondly, it would be beneficial to consider linear colliders with polarised beams. It is well understood that in such scenarios, one could notably enhance $S/B$ and $S/\sqrt{B}$ ratios for a variety of observables, thus exploring new effects. For instance, see \cite{Aguilar-Saavedra:2024hwd} for the proposal on studying post-decay entanglement, that is, entanglement of a top quark with the $Wb$ decay products of the other top. Thirdly, a comprehensive investigation of the potential of a muon collider operating between 3 and 10 TeV would be highly valuable. In this energy range, the primary production mechanisms for $t \bar t$ pairs would likely involve $\gamma\gamma$ and $WW$ fusion, enabling a much broader range of top-quark couplings and phenomena to be explored.

\bigskip

\section*{Acknowledgements}
We are greatful to Gauthier Durieux and Rafael Aoude for comments on the manuscript, and to Marco Zaro and all developers of {\tt MadGraph5\_aMC@NLO}, for their support and help throughout the years. 

FM acknowledges support by FRS-FNRS (Belgian National Scientific Research Fund) IISN projects 4.4503.16. CS and EV are supported by the European Union’s Horizon 2020 research and innovation programme under the EFT4NP project (grant agreement no.\@ 949451) and by a Royal Society University Research Fellowship through grant URF/ R1/ 201553. ST is supported by a FRIA (Fonds pour la formation \`a la Recherche dans l’Industrie et dans l’Agriculture) Grant of the FRS-FNRS (Belgian National Scientific Research Fund).

\clearpage

\appendix

\section{Common factors for spin correlation coefficients} \label{app:F}

In this Appendix we collect the expressions for the various common factors $F^{[i]}$ appearing in the spin correlation coefficients $A$ and $\widetilde{C}_{ij}$. \\

For SM top pair production, they are given by:
\begin{align}
    F^{[ 0]} &= 48 e^4 \Big(\Qt^2 \Ql^2 + 2 \re \frac{ 4 \Qt \Ql \gVt \gVl \mt^2 }{\DZ} + \frac{16 \gVt^2 \mt^4 (\gVl^2 + \gAl^2)}{|\DZ|^2} \Big), \label{F0}\\
    F^{[ 1]} &= 192 e^4 \gAt \gAl \mt^2 \beta \Big(\frac{16 \gVt \gVl \mt^2 }{|\DZ|^2} + 2 \re \frac{\Qt \Ql }{\DZ} \Big), \label{F1}\\
    F^{[ 2]} &= \frac{768 e^4 \gAt^2 \mt^4 \beta^2 (\gVl^2 + \gAl^2)}{|\DZ|^2}, \label{F2}
\end{align}
where $\DZ = \cW^2 \sW^2 (4 \mt^2-(1-\beta^2) \, \mz^2 )$ is the propagator denominator for the $Z$ boson, together with the electroweak angles coming from the $Z$ couplings on either side of it. The $Z$ width may be implemented by replacing $\mz^2 \to \mz^2 - i \mz \Gamma_Z$ in $\DZ$. 

The $F^{[0]}$, $F^{([1]}$ and $F^{(2)}$ coefficients already appeared in \cite{Maltoni:2024tul} for the process $q \bar q \to Z/\gamma \to t \bar t$, our expressions for \eqref{F0}--\eqref{F2} are consistent with those therein, up to the different color factors. \medskip

For SMEFT top pair production at order $1/\Lambda^2$, the common factors for interference at order $\Qt$ and $\gVt$ are given by:
\begin{align}
    F^{[6,0,V]} &= \frac{96 e^2 \mt^2}{\Lambda^2} \re \Big( c_\pV \frac{e^2 \Qt \Ql \gVl v^2}{\DZ} + c_\pV \frac{4 e^2 \gVt \mt^2 v^2 (\gAl^2 + \gVl^2)}{|\DZ|^2} \nonumber \\    
    &\quad +c_{\VV} \frac{2 \Ql \Qt}{1 - \beta^2} + (c_{\VV} \gVl - c_{\VA} \gAl) \frac{8  \gVt \mt^2}{\DZ (1-\beta^2)}   
    \Big) , \label{F60V}\\
    F^{[6,0,A]} &= \frac{96 e^2 \mt^2 \beta}{\Lambda^2} \re \Big(- c_\pA \frac{e^2 \Qt \Ql \gAl v^2}{\DZ} - c_\pA \frac{8 e^2 \gVt \mt^2 v^2 \gAl \gVl}{|\DZ|^2} \nonumber \\
    &\quad + c_{\AA} \frac{2 \Ql \Qt}{1-\beta^2} + (c_{\AA} \gVl - c_{\AV} \gAl) \frac{8 \gVt \mt^2}{\DZ (1-\beta^2)} \Big), \label{F60A}\\
    F^{[6,0,\text{D}]} &= \frac{192 \sqrt{2} \, e^3 \mt v}{\Lambda^2} \re \Big( c_\tA \Ql^2 \Qt - c_\tZ \frac{16 \cW \sW \gVt (\gAl^2 + \gVl^2) \mt^4}{|\DZ|^2} \nonumber \\
    &\quad + \big( c_\tA \gVt - c_\tZ \Qt \cW \sW \big) \frac{4 \gVl \mt^2 \Ql}{\DZ} \Big), \label{F60D}
\end{align}
for the vector \eqref{cV}, axial \eqref{cA}, and dipole \eqref{cD} operators, respectively. We have denoted $v$ the Higgs vev, not to be confused with the top velocity, $\beta$. The same common factors for interference at order $\gAt$ are given by:
\begin{align}
    F^{[6,1,V]} &= \frac{768 e^2 \mt^4 \gAt \beta}{\Lambda^2} \re \Big( c_\pV \frac{e^2 \gVl \gAl v^2 }{|\DZ|^2} + \frac{c_{\VV} \gAl - c_{\VA} \gVl}{\DZ (1-\beta^2)}\Big), \label{F61V} \\
    F^{[6,1,A]} &= \frac{768 e^2 \mt^4 \gAt \beta^2}{\Lambda^2} \re \Big( -c_\pA \frac{e^2 (\gAl^2 + \gVl^2) v^2}{2|\DZ|^2} + \frac{c_{\AA} \gAl - c_{\AV} \gVl}{\DZ(1-\beta^2)} \Big), \label{F61A} \\
    F^{[6,1,\text{D}]} &= \frac{768 \sqrt{2} \, e^3 \mt^3 v \gAt \gAl \beta}{\Lambda^2} \re \Big( c_\tA \frac{\Ql}{\DZ} - c_\tZ \frac{8 \cW \sW \gVl \mt^2}{|\DZ|^2} \Big) \label{F61D}.
\end{align}
At order $1/\Lambda^4$, the common factors for the purely vector, purely axial, and purely dipole terms are:
\begin{align}
    F^{[8,\VV]} &= \frac{192 \mt^4}{\Lambda^4} \Big( c_\pV^2 \frac{e^4 (\gAl^2 + \gVl^2) v^4}{|\DZ|^2} +  \frac{4(c_{\VV}^2 + c_{\VA}^2)}{(1-\beta^2)^2} \nonumber \\
    &\quad + 2  \re \frac{2 c_\pV (c_{\VV} \gVl - c_{\VA} \gAl) e^2 v^2}{\DZ (1-\beta^2)} \Big), \label{F8VV}\\
    F^{[8,\AA]} &= \frac{192 \mt^4 \beta^2}{\Lambda^4} \Big(  c_\pA^2 \frac{e^2  (\gAl^2 + \gVl^2) v^4}{|\DZ|^2} + \frac{4(c_{\AA}^2+c_{\AV}^2)}{(1-\beta^2)^2} \nonumber \\
    &\quad - 2  \re \frac{2 c_\pA(c_{\AA} \gAl - c_{\AV} \gVl)e^2 v^2}{\DZ (1-\beta^2)} \Big), \label{F8AA}\\
    F^{[8,\DD]} &= \frac{384 \mt^2 v^2 e^2}{\Lambda^4} \Big( c_\tA^2 \frac{\Ql^2}{1-\beta^2} + c_\tZ^2 \frac{16 \mt^4 \cW^2 \sW^2 (\gAl^2 + \gVl^2)}{|\DZ|^2 (1-\beta^2)}\nonumber \\
    &\quad - 2 \re \frac{4 c_\tA c_\tZ \Ql \gVl \mt^2}{\DZ(1-\beta^2)} \Big). \label{F8DD}
\end{align}

The factors for the mixed vector--axial, vector--dipole, and axial--dipole terms are given by:
\begin{align}
    F^{[8,\VA]} &= \frac{384 \mt^4 \beta}{\Lambda^4} \re \Big(- c_\pV c_\pA \frac{e^4 \gAl\gVl v^4}{|\DZ|^2} + \frac{2(c_\AA c_\VV + c_\AV c_\VA)}{(1-\beta^2)^2} \nonumber \\
    &\quad -(c_\VV c_\pA \gAl - c_\AA c_\pV \gVl + c_\AV c_\pV \gAl - c_\VA c_\pA \gVl) \frac{e^2 v^2}{\DZ (1-\beta^2)} \Big), \label{F8VA}\\
    F^{[8,\VD]} &= \frac{384 \sqrt{2} \, e \, \mt^3 v}{\Lambda^4} \re \Big(c_\tA c_\VV \frac{2 \Ql}{1-\beta^2} - c_\pV c_\tZ \frac{4 \cW \sW e^2 (\gAl^2+\gVl^2) \mt^2 v^2}{|\DZ|^2} \nonumber \\
    &\quad + c_\pV c_\tA \frac{e^2 \gVl \Ql v^2}{\DZ} - c_\tZ (c_\VV \gVl - c_\VA \gAl) \frac{8 \cW \sW \mt^2}{\DZ (1-\beta^2)} \Big), \label{F8VD}\\
    F^{[8,\AD]} &= \frac{384 \sqrt{2} \, e \, \mt^3 v \beta}{\Lambda^4} \re \Big( c_\tA c_\AA \frac{2 \Ql}{1-\beta^2} + c_\pA c_\tZ \frac{8 \cW \sW e^2 \gAl \gVl \mt^2 v^2}{|\DZ|^2}\nonumber \\
    &\quad - c_\pA c_\tA \frac{e^2 \gAl \Ql v^2}{\DZ} - c_\tZ (c_\AA \gVl - c_\AV \gAl) \frac{8 \cW \sW \mt^2}{\DZ(1-\beta^2)}\Big). \label{F8AD}
\end{align}

\clearpage

\bibliographystyle{JHEP}
\bibliography{bibfr.bib}

\providecommand{\href}[2]{#2}\begingroup\raggedright\begin{thebibliography}{100}

\bibitem{CDF:1994juo}
{\scshape CDF} collaboration, \emph{{Evidence for top quark production in
  $\bar{p}p$ collisions at $\sqrt{s} = 1.8$ TeV}},
  \href{https://doi.org/10.1103/PhysRevLett.73.225}{\emph{Phys. Rev. Lett.}
  {\bfseries 73} (1994) 225}
  [\href{https://arxiv.org/abs/hep-ex/9405005}{{\ttfamily hep-ex/9405005}}].

\bibitem{ParticleDataGroup:2022pth}
{\scshape Particle Data Group} collaboration, \emph{{Review of Particle
  Physics}}, \href{https://doi.org/10.1093/ptep/ptac097}{\emph{PTEP} {\bfseries
  2022} (2022) 083C01}.

\bibitem{CMS:2023wnd}
{\scshape CMS} collaboration, \emph{{Combination of measurements of the top
  quark mass from data collected by the ATLAS and CMS experiments at
  $\sqrt{s}=7$ and $8~\mathrm{TeV}$}},
  \href{https://arxiv.org/abs/2402.08713}{{\ttfamily 2402.08713}}.

\bibitem{Beneke:2015kwa}
M.~Beneke, Y.~Kiyo, P.~Marquard, A.~Penin, J.~Piclum and M.~Steinhauser,
  \emph{{Next-to-Next-to-Next-to-Leading Order QCD Prediction for the Top
  Antitop $S$-Wave Pair Production Cross Section Near Threshold in $e^+e^-$
  Annihilation}},
  \href{https://doi.org/10.1103/PhysRevLett.115.192001}{\emph{Phys. Rev. Lett.}
  {\bfseries 115} (2015) 192001}
  [\href{https://arxiv.org/abs/1506.06864}{{\ttfamily 1506.06864}}].

\bibitem{Behnke:2013xla}
T.~Behnke, J.E.~Brau, B.~Foster, J.~Fuster, M.~Harrison, J.M.~Paterson et~al.,
  eds., \emph{{The International Linear Collider Technical Design Report -
  Volume 1: Executive Summary}},
  \href{https://arxiv.org/abs/1306.6327}{{\ttfamily 1306.6327}}.

\bibitem{CLICdp:2018esa}
{\scshape CLICdp} collaboration, \emph{{Top-Quark Physics at the CLIC
  Electron-Positron Linear Collider}},
  \href{https://doi.org/10.1007/JHEP11(2019)003}{\emph{JHEP} {\bfseries 11}
  (2019) 003} [\href{https://arxiv.org/abs/1807.02441}{{\ttfamily
  1807.02441}}].

\bibitem{FCC:2018byv}
{\scshape FCC} collaboration, \emph{{FCC Physics Opportunities}: {Future
  Circular Collider Conceptual Design Report Volume 1}},
  \href{https://doi.org/10.1140/epjc/s10052-019-6904-3}{\emph{Eur. Phys. J. C}
  {\bfseries 79} (2019) 474}.

\bibitem{Janot:2015mqv}
{\scshape FCC Design Study Group} collaboration, \emph{{Precision measurements
  of the top quark couplings at the FCC}},
  \href{https://doi.org/10.22323/1.234.0333}{\emph{PoS} {\bfseries EPS-HEP2015}
  (2015) 333} [\href{https://arxiv.org/abs/1510.09056}{{\ttfamily
  1510.09056}}].

\bibitem{Durieux:2018ekg}
G.~Durieux and O.~Matsedonskyi, \emph{{The top-quark window on compositeness at
  future lepton colliders}},
  \href{https://doi.org/10.1007/JHEP01(2019)072}{\emph{JHEP} {\bfseries 01}
  (2019) 072} [\href{https://arxiv.org/abs/1807.10273}{{\ttfamily
  1807.10273}}].

\bibitem{Vryonidou:2018eyv}
E.~Vryonidou and C.~Zhang, \emph{{Dimension-six electroweak top-loop effects in
  Higgs production and decay}},
  \href{https://doi.org/10.1007/JHEP08(2018)036}{\emph{JHEP} {\bfseries 08}
  (2018) 036} [\href{https://arxiv.org/abs/1804.09766}{{\ttfamily
  1804.09766}}].

\bibitem{Durieux:2018ggn}
G.~Durieux, J.~Gu, E.~Vryonidou and C.~Zhang, \emph{{Probing top-quark
  couplings indirectly at Higgs factories}},
  \href{https://doi.org/10.1088/1674-1137/42/12/123107}{\emph{Chin. Phys. C}
  {\bfseries 42} (2018) 123107}
  [\href{https://arxiv.org/abs/1809.03520}{{\ttfamily 1809.03520}}].

\bibitem{Durieux:2018tev}
G.~Durieux, M.~Perell\'o, M.~Vos and C.~Zhang, \emph{{Global and optimal probes
  for the top-quark effective field theory at future lepton colliders}},
  \href{https://doi.org/10.1007/JHEP10(2018)168}{\emph{JHEP} {\bfseries 10}
  (2018) 168} [\href{https://arxiv.org/abs/1807.02121}{{\ttfamily
  1807.02121}}].

\bibitem{Durieux:2019rbz}
G.~Durieux, A.~Irles, V.~Miralles, A.~Pe\~nuelas, R.~P\"oschl, M.~Perell\'o
  et~al., \emph{{The electro-weak couplings of the top and bottom quarks
  \textemdash{} Global fit and future prospects}},
  \href{https://doi.org/10.1007/JHEP12(2019)098}{\emph{JHEP} {\bfseries 12}
  (2019) 98} [\href{https://arxiv.org/abs/1907.10619}{{\ttfamily 1907.10619}}].

\bibitem{Jung:2020uzh}
S.~Jung, J.~Lee, M.~Perell\'o, J.~Tian and M.~Vos, \emph{{Higgs, top quark, and
  electroweak precision measurements at future e+e- colliders: A combined
  effective field theory analysis with renormalization mixing}},
  \href{https://doi.org/10.1103/PhysRevD.105.016003}{\emph{Phys. Rev. D}
  {\bfseries 105} (2022) 016003}
  [\href{https://arxiv.org/abs/2006.14631}{{\ttfamily 2006.14631}}].

\bibitem{Bernardi:2022hny}
G.~Bernardi et~al., \emph{{The Future Circular Collider: a Summary for the US
  2021 Snowmass Process}},  \href{https://arxiv.org/abs/2203.06520}{{\ttfamily
  2203.06520}}.

\bibitem{deBlas:2022ofj}
J.~de~Blas, Y.~Du, C.~Grojean, J.~Gu, V.~Miralles, M.E.~Peskin et~al.,
  \emph{{Global SMEFT Fits at Future Colliders}},  in \emph{{Snowmass 2021}},
  6, 2022 [\href{https://arxiv.org/abs/2206.08326}{{\ttfamily 2206.08326}}].

\bibitem{Durieux:2022cvf}
G.~Durieux, A.G.~Camacho, L.~Mantani, V.~Miralles, M.M.~L\'opez,
  M.~Ll\'acer~Moreno et~al., \emph{{Snowmass White Paper: prospects for the
  measurement of top-quark couplings}},  in \emph{{Snowmass 2021}}, 5, 2022
  [\href{https://arxiv.org/abs/2205.02140}{{\ttfamily 2205.02140}}].

\bibitem{Banelli:2020iau}
G.~Banelli, E.~Salvioni, J.~Serra, T.~Theil and A.~Weiler, \emph{{The Present
  and Future of Four Top Operators}},
  \href{https://doi.org/10.1007/JHEP02(2021)043}{\emph{JHEP} {\bfseries 02}
  (2021) 043} [\href{https://arxiv.org/abs/2010.05915}{{\ttfamily
  2010.05915}}].

\bibitem{PhysRev.103.1901}
K.~Lande, E.T.~Booth, J.~Impeduglia, L.M.~Lederman and W.~Chinowsky,
  \emph{Observation of long-lived neutral $v$ particles},
  \href{https://doi.org/10.1103/PhysRev.103.1901}{\emph{Phys. Rev.} {\bfseries
  103} (1956) 1901}.

\bibitem{BaBar:2007kib}
{\scshape BaBar} collaboration, \emph{{Evidence for $D^{\,0} -
  \overline{D}^{\,0}$ Mixing}},
  \href{https://doi.org/10.1103/PhysRevLett.98.211802}{\emph{Phys. Rev. Lett.}
  {\bfseries 98} (2007) 211802}
  [\href{https://arxiv.org/abs/hep-ex/0703020}{{\ttfamily hep-ex/0703020}}].

\bibitem{CDF:2007bdz}
{\scshape CDF} collaboration, \emph{{Evidence for $D^0 - \bar{D}^0$ mixing
  using the CDF II Detector}},
  \href{https://doi.org/10.1103/PhysRevLett.100.121802}{\emph{Phys. Rev. Lett.}
  {\bfseries 100} (2008) 121802}
  [\href{https://arxiv.org/abs/0712.1567}{{\ttfamily 0712.1567}}].

\bibitem{LHCb:2012zll}
{\scshape LHCb} collaboration, \emph{{Observation of $D^0 - \overline{D}^0$
  oscillations}},
  \href{https://doi.org/10.1103/PhysRevLett.110.101802}{\emph{Phys. Rev. Lett.}
  {\bfseries 110} (2013) 101802}
  [\href{https://arxiv.org/abs/1211.1230}{{\ttfamily 1211.1230}}].

\bibitem{ARGUS:1987xtv}
{\scshape ARGUS} collaboration, \emph{{Observation of B0 - anti-B0 Mixing}},
  \href{https://doi.org/10.1016/0370-2693(87)91177-4}{\emph{Phys. Lett. B}
  {\bfseries 192} (1987) 245}.

\bibitem{D0:2006oeb}
{\scshape D0} collaboration, \emph{{First direct two-sided bound on the
  $B^0_{s}$ oscillation frequency}},
  \href{https://doi.org/10.1103/PhysRevLett.97.021802}{\emph{Phys. Rev. Lett.}
  {\bfseries 97} (2006) 021802}
  [\href{https://arxiv.org/abs/hep-ex/0603029}{{\ttfamily hep-ex/0603029}}].

\bibitem{CDF:2006imy}
{\scshape CDF} collaboration, \emph{{Observation of $B^0_s - \bar{B}^0_s$
  Oscillations}},
  \href{https://doi.org/10.1103/PhysRevLett.97.242003}{\emph{Phys. Rev. Lett.}
  {\bfseries 97} (2006) 242003}
  [\href{https://arxiv.org/abs/hep-ex/0609040}{{\ttfamily hep-ex/0609040}}].

\bibitem{Belle:2002bwy}
{\scshape Belle} collaboration, \emph{{Observation of mixing induced CP
  violation in the neutral $B$ meson system}},
  \href{https://doi.org/10.1103/PhysRevD.66.032007}{\emph{Phys. Rev. D}
  {\bfseries 66} (2002) 032007}
  [\href{https://arxiv.org/abs/hep-ex/0202027}{{\ttfamily hep-ex/0202027}}].

\bibitem{LHCb:2013lrq}
{\scshape LHCb} collaboration, \emph{{Precision measurement of the
  $B^{0}_{s}$-$\bar{B}^{0}_{s}$ oscillation frequency with the decay
  $B^{0}_{s}\rightarrow D^{-}_{s}\pi^{+}$}},
  \href{https://doi.org/10.1088/1367-2630/15/5/053021}{\emph{New J. Phys.}
  {\bfseries 15} (2013) 053021}
  [\href{https://arxiv.org/abs/1304.4741}{{\ttfamily 1304.4741}}].

\bibitem{ATLAS:2014aus}
{\scshape ATLAS} collaboration, \emph{{Measurements of spin correlation in
  top-antitop quark events from proton-proton collisions at $\sqrt{s}=7$ TeV
  using the ATLAS detector}},
  \href{https://doi.org/10.1103/PhysRevD.90.112016}{\emph{Phys. Rev. D}
  {\bfseries 90} (2014) 112016}
  [\href{https://arxiv.org/abs/1407.4314}{{\ttfamily 1407.4314}}].

\bibitem{CMS:2015cal}
{\scshape CMS} collaboration, \emph{{Measurement of Spin Correlations in
  $t\bar{t}$ Production using the Matrix Element Method in the Muon+Jets Final
  State in $pp$ Collisions at $\sqrt{s} =$ 8 TeV}},
  \href{https://doi.org/10.1016/j.physletb.2016.05.005}{\emph{Phys. Lett. B}
  {\bfseries 758} (2016) 321}
  [\href{https://arxiv.org/abs/1511.06170}{{\ttfamily 1511.06170}}].

\bibitem{CMS:2016piu}
{\scshape CMS} collaboration, \emph{{Measurements of t t-bar spin correlations
  and top quark polarization using dilepton final states in pp collisions at
  sqrt(s) = 8 TeV}},
  \href{https://doi.org/10.1103/PhysRevD.93.052007}{\emph{Phys. Rev. D}
  {\bfseries 93} (2016) 052007}
  [\href{https://arxiv.org/abs/1601.01107}{{\ttfamily 1601.01107}}].

\bibitem{ATLAS:2016bac}
{\scshape ATLAS} collaboration, \emph{{Measurements of top quark spin
  observables in $ t\overline{t} $ events using dilepton final states in $
  \sqrt{s}=8 $ TeV pp collisions with the ATLAS detector}},
  \href{https://doi.org/10.1007/JHEP03(2017)113}{\emph{JHEP} {\bfseries 03}
  (2017) 113} [\href{https://arxiv.org/abs/1612.07004}{{\ttfamily
  1612.07004}}].

\bibitem{ATLAS:2019hau}
{\scshape ATLAS} collaboration, \emph{{Measurement of the $t\bar{t}$ production
  cross-section and lepton differential distributions in $e\mu $ dilepton
  events from $pp$ collisions at $\sqrt{s}=13\,\text {TeV}$ with the ATLAS
  detector}}, \href{https://doi.org/10.1140/epjc/s10052-020-7907-9}{\emph{Eur.
  Phys. J. C} {\bfseries 80} (2020) 528}
  [\href{https://arxiv.org/abs/1910.08819}{{\ttfamily 1910.08819}}].

\bibitem{ATLAS:2019zrq}
{\scshape ATLAS} collaboration, \emph{{Measurements of top-quark pair spin
  correlations in the $e\mu$ channel at $\sqrt{s} = 13$ TeV using $pp$
  collisions in the ATLAS detector}},
  \href{https://doi.org/10.1140/epjc/s10052-020-8181-6}{\emph{Eur. Phys. J. C}
  {\bfseries 80} (2020) 754}
  [\href{https://arxiv.org/abs/1903.07570}{{\ttfamily 1903.07570}}].

\bibitem{CMS:2019nrx}
{\scshape CMS} collaboration, \emph{{Measurement of the top quark polarization
  and $\mathrm{t\bar{t}}$ spin correlations using dilepton final states in
  proton-proton collisions at $\sqrt{s} =$ 13 TeV}},
  \href{https://doi.org/10.1103/PhysRevD.100.072002}{\emph{Phys. Rev. D}
  {\bfseries 100} (2019) 072002}
  [\href{https://arxiv.org/abs/1907.03729}{{\ttfamily 1907.03729}}].

\bibitem{ATLAS:2023fsd}
{\scshape ATLAS} collaboration, \emph{{Observation of quantum entanglement in
  top-quark pairs using the ATLAS detector}},
  \href{https://arxiv.org/abs/2311.07288}{{\ttfamily 2311.07288}}.

\bibitem{CMS:2024pts}
{\scshape CMS} collaboration, \emph{{Observation of quantum entanglement in top
  quark pair production in proton-proton collisions at $\sqrt{s}$ = 13 TeV}},
  \href{https://arxiv.org/abs/2406.03976}{{\ttfamily 2406.03976}}.

\bibitem{CMS:2024vqh}
{\scshape CMS} collaboration, \emph{{Measurements of polarization, spin
  correlations, and entanglement in top quark pairs using lepton+jets events
  from pp collisions at $\sqrt{s}=13~\mathrm{TeV}$}}, .

\bibitem{Maltoni:2024tul}
F.~Maltoni, C.~Severi, S.~Tentori and E.~Vryonidou, \emph{{Quantum detection of
  new physics in top-quark pair production at the LHC}},
  \href{https://doi.org/10.1007/JHEP03(2024)099}{\emph{JHEP} {\bfseries 03}
  (2024) 099} [\href{https://arxiv.org/abs/2401.08751}{{\ttfamily
  2401.08751}}].

\bibitem{Barr:2024djo}
A.J.~Barr, M.~Fabbrichesi, R.~Floreanini, E.~Gabrielli and L.~Marzola,
  \emph{{Quantum entanglement and Bell inequality violation at colliders}},
  \href{https://arxiv.org/abs/2402.07972}{{\ttfamily 2402.07972}}.

\bibitem{Aoude:2022imd}
R.~Aoude, E.~Madge, F.~Maltoni and L.~Mantani, \emph{{Quantum SMEFT tomography:
  Top quark pair production at the LHC}},
  \href{https://doi.org/10.1103/PhysRevD.106.055007}{\emph{Phys. Rev. D}
  {\bfseries 106} (2022) 055007}
  [\href{https://arxiv.org/abs/2203.05619}{{\ttfamily 2203.05619}}].

\bibitem{Severi:2022qjy}
C.~Severi and E.~Vryonidou, \emph{{Quantum entanglement and top spin
  correlations in SMEFT at higher orders}},
  \href{https://doi.org/10.1007/JHEP01(2023)148}{\emph{JHEP} {\bfseries 01}
  (2023) 148} [\href{https://arxiv.org/abs/2210.09330}{{\ttfamily
  2210.09330}}].

\bibitem{Fabbrichesi:2024xtq}
M.~Fabbrichesi and L.~Marzola, \emph{{Dipole momenta and compositeness of the
  $\tau$ lepton at Belle II}},
  \href{https://arxiv.org/abs/2401.04449}{{\ttfamily 2401.04449}}.

\bibitem{Fabbrichesi:2022ovb}
M.~Fabbrichesi, R.~Floreanini and E.~Gabrielli, \emph{{Constraining new physics
  in entangled two-qubit systems: top-quark, tau-lepton and photon pairs}},
  \href{https://doi.org/10.1140/epjc/s10052-023-11307-2}{\emph{Eur. Phys. J. C}
  {\bfseries 83} (2023) 162}
  [\href{https://arxiv.org/abs/2208.11723}{{\ttfamily 2208.11723}}].

\bibitem{Altakach:2022ywa}
M.M.~Altakach, P.~Lamba, F.~Maltoni, K.~Mawatari and K.~Sakurai, \emph{{Quantum
  information and CP measurement in
  H\textrightarrow{}\ensuremath{\tau}+\ensuremath{\tau}- at future lepton
  colliders}}, \href{https://doi.org/10.1103/PhysRevD.107.093002}{\emph{Phys.
  Rev. D} {\bfseries 107} (2023) 093002}
  [\href{https://arxiv.org/abs/2211.10513}{{\ttfamily 2211.10513}}].

\bibitem{Aoude:2023hxv}
R.~Aoude, E.~Madge, F.~Maltoni and L.~Mantani, \emph{{Probing new physics
  through entanglement in diboson production}},
  \href{https://doi.org/10.1007/JHEP12(2023)017}{\emph{JHEP} {\bfseries 12}
  (2023) 017} [\href{https://arxiv.org/abs/2307.09675}{{\ttfamily
  2307.09675}}].

\bibitem{Bernal:2023ruk}
A.~Bernal, P.~Caban and J.~Rembieli\'nski, \emph{{Entanglement and Bell
  inequalities violation in $H\rightarrow ZZ$ with anomalous coupling}},
  \href{https://doi.org/10.1140/epjc/s10052-023-12216-0}{\emph{Eur. Phys. J. C}
  {\bfseries 83} (2023) 1050}
  [\href{https://arxiv.org/abs/2307.13496}{{\ttfamily 2307.13496}}].

\bibitem{Fabbrichesi:2023jep}
M.~Fabbrichesi, R.~Floreanini, E.~Gabrielli and L.~Marzola, \emph{{Stringent
  bounds on HWW and HZZ anomalous couplings with quantum tomography at the
  LHC}}, \href{https://doi.org/10.1007/JHEP09(2023)195}{\emph{JHEP} {\bfseries
  09} (2023) 195} [\href{https://arxiv.org/abs/2304.02403}{{\ttfamily
  2304.02403}}].

\bibitem{Aebischer:2018csl}
J.~Aebischer, C.~Bobeth, A.J.~Buras and D.M.~Straub, \emph{{Anatomy of
  $\varepsilon '/\varepsilon $ beyond the standard model}},
  \href{https://doi.org/10.1140/epjc/s10052-019-6715-6}{\emph{Eur. Phys. J. C}
  {\bfseries 79} (2019) 219}
  [\href{https://arxiv.org/abs/1808.00466}{{\ttfamily 1808.00466}}].

\bibitem{Charles:2020dfl}
J.~Charles, S.~Descotes-Genon, Z.~Ligeti, S.~Monteil, M.~Papucci, K.~Trabelsi
  et~al., \emph{{New physics in $B$ meson mixing: future sensitivity and
  limitations}}, \href{https://doi.org/10.1103/PhysRevD.102.056023}{\emph{Phys.
  Rev. D} {\bfseries 102} (2020) 056023}
  [\href{https://arxiv.org/abs/2006.04824}{{\ttfamily 2006.04824}}].

\bibitem{Falkowski:2023hsg}
A.~Falkowski, \emph{{Lectures on SMEFT}},
  \href{https://doi.org/10.1140/epjc/s10052-023-11821-3}{\emph{Eur. Phys. J. C}
  {\bfseries 83} (2023) 656}.

\bibitem{Aguilar-Saavedra:2018ksv}
D.~Barducci et~al., \emph{{Interpreting top-quark LHC measurements in the
  standard-model effective field theory}},
  \href{https://arxiv.org/abs/1802.07237}{{\ttfamily 1802.07237}}.

\bibitem{Hartland:2019bjb}
N.P.~Hartland, F.~Maltoni, E.R.~Nocera, J.~Rojo, E.~Slade, E.~Vryonidou et~al.,
  \emph{{A Monte Carlo global analysis of the Standard Model Effective Field
  Theory: the top quark sector}},
  \href{https://doi.org/10.1007/JHEP04(2019)100}{\emph{JHEP} {\bfseries 04}
  (2019) 100} [\href{https://arxiv.org/abs/1901.05965}{{\ttfamily
  1901.05965}}].

\bibitem{Bissmann:2019gfc}
S.~Bi\ss{}mann, J.~Erdmann, C.~Grunwald, G.~Hiller and K.~Kr\"oninger,
  \emph{{Constraining top-quark couplings combining top-quark and
  $\boldsymbol{B}$ decay observables}},
  \href{https://doi.org/10.1140/epjc/s10052-020-7680-9}{\emph{Eur. Phys. J. C}
  {\bfseries 80} (2020) 136}
  [\href{https://arxiv.org/abs/1909.13632}{{\ttfamily 1909.13632}}].

\bibitem{Brivio:2019ius}
I.~Brivio, S.~Bruggisser, F.~Maltoni, R.~Moutafis, T.~Plehn, E.~Vryonidou
  et~al., \emph{{O new physics, where art thou? A global search in the top
  sector}}, \href{https://doi.org/10.1007/JHEP02(2020)131}{\emph{JHEP}
  {\bfseries 02} (2020) 131}
  [\href{https://arxiv.org/abs/1910.03606}{{\ttfamily 1910.03606}}].

\bibitem{Ellis:2020unq}
J.~Ellis, M.~Madigan, K.~Mimasu, V.~Sanz and T.~You, \emph{{Top, Higgs, Diboson
  and Electroweak Fit to the Standard Model Effective Field Theory}},
  \href{https://doi.org/10.1007/JHEP04(2021)279}{\emph{JHEP} {\bfseries 04}
  (2021) 279} [\href{https://arxiv.org/abs/2012.02779}{{\ttfamily
  2012.02779}}].

\bibitem{Bissmann:2020mfi}
S.~Bi\ss{}mann, C.~Grunwald, G.~Hiller and K.~Kr\"oninger, \emph{{Top and
  Beauty synergies in SMEFT-fits at present and future colliders}},
  \href{https://doi.org/10.1007/JHEP06(2021)010}{\emph{JHEP} {\bfseries 06}
  (2021) 010} [\href{https://arxiv.org/abs/2012.10456}{{\ttfamily
  2012.10456}}].

\bibitem{Ethier:2021bye}
{\scshape SMEFiT} collaboration, \emph{{Combined SMEFT interpretation of Higgs,
  diboson, and top quark data from the LHC}},
  \href{https://doi.org/10.1007/JHEP11(2021)089}{\emph{JHEP} {\bfseries 11}
  (2021) 089} [\href{https://arxiv.org/abs/2105.00006}{{\ttfamily
  2105.00006}}].

\bibitem{Miralles:2021dyw}
V.~Miralles, M.M.~L\'opez, M.M.~Ll\'acer, A.~Pe\~nuelas, M.~Perell\'o and
  M.~Vos, \emph{{The top quark electro-weak couplings after LHC Run 2}},
  \href{https://doi.org/10.1007/JHEP02(2022)032}{\emph{JHEP} {\bfseries 02}
  (2022) 032} [\href{https://arxiv.org/abs/2107.13917}{{\ttfamily
  2107.13917}}].

\bibitem{Liu:2022vgo}
Y.~Liu, Y.~Wang, C.~Zhang, L.~Zhang and J.~Gu, \emph{{Probing top-quark
  operators with precision electroweak measurements*}},
  \href{https://doi.org/10.1088/1674-1137/ac82e1}{\emph{Chin. Phys. C}
  {\bfseries 46} (2022) 113105}
  [\href{https://arxiv.org/abs/2205.05655}{{\ttfamily 2205.05655}}].

\bibitem{Kassabov:2023hbm}
Z.~Kassabov, M.~Madigan, L.~Mantani, J.~Moore, M.~Morales~Alvarado, J.~Rojo
  et~al., \emph{{The top quark legacy of the LHC Run II for PDF and SMEFT
  analyses}}, \href{https://doi.org/10.1007/JHEP05(2023)205}{\emph{JHEP}
  {\bfseries 05} (2023) 205}
  [\href{https://arxiv.org/abs/2303.06159}{{\ttfamily 2303.06159}}].

\bibitem{Allwicher:2023shc}
L.~Allwicher, C.~Cornella, G.~Isidori and B.A.~Stefanek, \emph{{New Physics in
  the Third Generation: A Comprehensive SMEFT Analysis and Future Prospects}},
  \href{https://arxiv.org/abs/2311.00020}{{\ttfamily 2311.00020}}.

\bibitem{Elmer:2023wtr}
N.~Elmer, M.~Madigan, T.~Plehn and N.~Schmal, \emph{{Staying on Top of
  SMEFT-Likelihood Analyses}},
  \href{https://arxiv.org/abs/2312.12502}{{\ttfamily 2312.12502}}.

\bibitem{Celada:2024mcf}
E.~Celada, T.~Giani, J.~ter Hoeve, L.~Mantani, J.~Rojo, A.N.~Rossia et~al.,
  \emph{{Mapping the SMEFT at High-Energy Colliders: from LEP and the (HL-)LHC
  to the FCC-ee}},  \href{https://arxiv.org/abs/2404.12809}{{\ttfamily
  2404.12809}}.

\bibitem{Wootters:1997id}
W.K.~Wootters, \emph{{Entanglement of formation of an arbitrary state of two
  qubits}}, \href{https://doi.org/10.1103/PhysRevLett.80.2245}{\emph{Phys. Rev.
  Lett.} {\bfseries 80} (1998) 2245}
  [\href{https://arxiv.org/abs/quant-ph/9709029}{{\ttfamily
  quant-ph/9709029}}].

\bibitem{Severi:2021cnj}
C.~Severi, C.D.E.~Boschi, F.~Maltoni and M.~Sioli, \emph{{Quantum tops at the
  LHC: from entanglement to Bell inequalities}},
  \href{https://doi.org/10.1140/epjc/s10052-022-10245-9}{\emph{Eur. Phys. J. C}
  {\bfseries 82} (2022) 285}
  [\href{https://arxiv.org/abs/2110.10112}{{\ttfamily 2110.10112}}].

\bibitem{2203.05582}
Y.~Afik and J.R.M.n.~de~Nova, \emph{{Quantum information with top quarks in
  QCD}}, \href{https://doi.org/10.22331/q-2022-09-29-820}{\emph{Quantum}
  {\bfseries 6} (2022) 820} [\href{https://arxiv.org/abs/2203.05582}{{\ttfamily
  2203.05582}}].

\bibitem{2205.00542}
J.A.~Aguilar-Saavedra and J.A.~Casas, \emph{{Improved tests of entanglement and
  Bell inequalities with LHC tops}},
  \href{https://doi.org/10.1140/epjc/s10052-022-10630-4}{\emph{Eur. Phys. J. C}
  {\bfseries 82} (2022) 666}
  [\href{https://arxiv.org/abs/2205.00542}{{\ttfamily 2205.00542}}].

\bibitem{2003.02280}
Y.~Afik and J.R.M.n.~de~Nova, \emph{{Entanglement and quantum tomography with
  top quarks at the LHC}},
  \href{https://doi.org/10.1140/epjp/s13360-021-01902-1}{\emph{Eur. Phys. J.
  Plus} {\bfseries 136} (2021) 907}
  [\href{https://arxiv.org/abs/2003.02280}{{\ttfamily 2003.02280}}].

\bibitem{Czakon:2020qbd}
M.~Czakon, A.~Mitov and R.~Poncelet, \emph{{NNLO QCD corrections to leptonic
  observables in top-quark pair production and decay}},
  \href{https://doi.org/10.1007/JHEP05(2021)212}{\emph{JHEP} {\bfseries 05}
  (2021) 212} [\href{https://arxiv.org/abs/2008.11133}{{\ttfamily
  2008.11133}}].

\bibitem{Mahlon:1997uc}
G.~Mahlon and S.J.~Parke, \emph{{Maximizing spin correlations in top quark pair
  production at the Tevatron}},
  \href{https://doi.org/10.1016/S0370-2693(97)00987-8}{\emph{Phys. Lett. B}
  {\bfseries 411} (1997) 173}
  [\href{https://arxiv.org/abs/hep-ph/9706304}{{\ttfamily hep-ph/9706304}}].

\bibitem{Parke:1996pr}
S.J.~Parke and Y.~Shadmi, \emph{{Spin correlations in top quark pair production
  at $e^{+} e^{-}$ colliders}},
  \href{https://doi.org/10.1016/0370-2693(96)00998-7}{\emph{Phys. Lett. B}
  {\bfseries 387} (1996) 199}
  [\href{https://arxiv.org/abs/hep-ph/9606419}{{\ttfamily hep-ph/9606419}}].

\bibitem{Fabbrichesi:2021npl}
M.~Fabbrichesi, R.~Floreanini and G.~Panizzo, \emph{{Testing Bell Inequalities
  at the LHC with Top-Quark Pairs}},
  \href{https://doi.org/10.1103/PhysRevLett.127.161801}{\emph{Phys. Rev. Lett.}
  {\bfseries 127} (2021) 161801}
  [\href{https://arxiv.org/abs/2102.11883}{{\ttfamily 2102.11883}}].

\bibitem{2209.03969}
Y.~Afik and J.R.M.n.~de~Nova, \emph{{Quantum Discord and Steering in Top Quarks
  at the LHC}},
  \href{https://doi.org/10.1103/PhysRevLett.130.221801}{\emph{Phys. Rev. Lett.}
  {\bfseries 130} (2023) 221801}
  [\href{https://arxiv.org/abs/2209.03969}{{\ttfamily 2209.03969}}].

\bibitem{Han:2023fci}
T.~Han, M.~Low and T.A.~Wu, \emph{{Quantum Entanglement and Bell Inequality
  Violation in Semi-Leptonic Top Decays}},
  \href{https://arxiv.org/abs/2310.17696}{{\ttfamily 2310.17696}}.

\bibitem{NNPDF:2021njg}
{\scshape NNPDF} collaboration, \emph{{The path to proton structure at 1\%
  accuracy}}, \href{https://doi.org/10.1140/epjc/s10052-022-10328-7}{\emph{Eur.
  Phys. J. C} {\bfseries 82} (2022) 428}
  [\href{https://arxiv.org/abs/2109.02653}{{\ttfamily 2109.02653}}].

\bibitem{Bernreuther:2015yna}
W.~Bernreuther, D.~Heisler and Z.-G.~Si, \emph{{A set of top quark spin
  correlation and polarization observables for the LHC: Standard Model
  predictions and new physics contributions}},
  \href{https://doi.org/10.1007/JHEP12(2015)026}{\emph{JHEP} {\bfseries 12}
  (2015) 026} [\href{https://arxiv.org/abs/1508.05271}{{\ttfamily
  1508.05271}}].

\bibitem{Dong:2023xiw}
Z.~Dong, D.~Gon\c{c}alves, K.~Kong and A.~Navarro, \emph{{When the Machine
  Chimes the Bell: Entanglement and Bell Inequalities with Boosted
  $t\bar{t}$}},  \href{https://arxiv.org/abs/2305.07075}{{\ttfamily
  2305.07075}}.

\bibitem{Hoang:2001mm}
A.H.~Hoang, A.V.~Manohar, I.W.~Stewart and T.~Teubner, \emph{{The Threshold t
  anti-t cross-section at NNLL order}},
  \href{https://doi.org/10.1103/PhysRevD.65.014014}{\emph{Phys. Rev. D}
  {\bfseries 65} (2002) 014014}
  [\href{https://arxiv.org/abs/hep-ph/0107144}{{\ttfamily hep-ph/0107144}}].

\bibitem{Hoang:2013uda}
A.H.~Hoang and M.~Stahlhofen, \emph{{The Top-Antitop Threshold at the ILC: NNLL
  QCD Uncertainties}},
  \href{https://doi.org/10.1007/JHEP05(2014)121}{\emph{JHEP} {\bfseries 05}
  (2014) 121} [\href{https://arxiv.org/abs/1309.6323}{{\ttfamily 1309.6323}}].

\bibitem{Bach:2017ggt}
F.~Bach, B.C.~Nejad, A.~Hoang, W.~Kilian, J.~Reuter, M.~Stahlhofen et~al.,
  \emph{{Fully-differential Top-Pair Production at a Lepton Collider: From
  Threshold to Continuum}},
  \href{https://doi.org/10.1007/JHEP03(2018)184}{\emph{JHEP} {\bfseries 03}
  (2018) 184} [\href{https://arxiv.org/abs/1712.02220}{{\ttfamily
  1712.02220}}].

\bibitem{Kiyo:2008bv}
Y.~Kiyo, J.H.~Kuhn, S.~Moch, M.~Steinhauser and P.~Uwer, \emph{{Top-quark pair
  production near threshold at LHC}},
  \href{https://doi.org/10.1140/epjc/s10052-009-0892-7}{\emph{Eur. Phys. J. C}
  {\bfseries 60} (2009) 375} [\href{https://arxiv.org/abs/0812.0919}{{\ttfamily
  0812.0919}}].

\bibitem{Aoude:2020mlg}
R.~Aoude, M.-Z.~Chung, Y.-t.~Huang, C.S.~Machado and M.-K.~Tam, \emph{{Silence
  of Binary Kerr Black Holes}},
  \href{https://doi.org/10.1103/PhysRevLett.125.181602}{\emph{Phys. Rev. Lett.}
  {\bfseries 125} (2020) 181602}
  [\href{https://arxiv.org/abs/2007.09486}{{\ttfamily 2007.09486}}].

\bibitem{Beane:2018oxh}
S.R.~Beane, D.B.~Kaplan, N.~Klco and M.J.~Savage, \emph{{Entanglement
  Suppression and Emergent Symmetries of Strong Interactions}},
  \href{https://doi.org/10.1103/PhysRevLett.122.102001}{\emph{Phys. Rev. Lett.}
  {\bfseries 122} (2019) 102001}
  [\href{https://arxiv.org/abs/1812.03138}{{\ttfamily 1812.03138}}].

\bibitem{Carena:2023vjc}
M.~Carena, I.~Low, C.E.M.~Wagner and M.-L.~Xiao, \emph{{Entanglement
  Suppression, Enhanced Symmetry and a Standard-Model-like Higgs Boson}},
  \href{https://arxiv.org/abs/2307.08112}{{\ttfamily 2307.08112}}.

\bibitem{Cervera-Lierta:2017tdt}
A.~Cervera-Lierta, J.I.~Latorre, J.~Rojo and L.~Rottoli, \emph{{Maximal
  Entanglement in High Energy Physics}},
  \href{https://doi.org/10.21468/SciPostPhys.3.5.036}{\emph{SciPost Phys.}
  {\bfseries 3} (2017) 036} [\href{https://arxiv.org/abs/1703.02989}{{\ttfamily
  1703.02989}}].

\bibitem{Low:2021ufv}
I.~Low and T.~Mehen, \emph{{Symmetry from entanglement suppression}},
  \href{https://doi.org/10.1103/PhysRevD.104.074014}{\emph{Phys. Rev. D}
  {\bfseries 104} (2021) 074014}
  [\href{https://arxiv.org/abs/2104.10835}{{\ttfamily 2104.10835}}].

\bibitem{Degrande:2020evl}
C.~Degrande, G.~Durieux, F.~Maltoni, K.~Mimasu, E.~Vryonidou and C.~Zhang,
  \emph{{Automated one-loop computations in the standard model effective field
  theory}}, \href{https://doi.org/10.1103/PhysRevD.103.096024}{\emph{Phys. Rev.
  D} {\bfseries 103} (2021) 096024}
  [\href{https://arxiv.org/abs/2008.11743}{{\ttfamily 2008.11743}}].

\bibitem{FCC:2018evy}
{\scshape FCC} collaboration, \emph{{FCC-ee: The Lepton Collider}: {Future
  Circular Collider Conceptual Design Report Volume 2}},
  \href{https://doi.org/10.1140/epjst/e2019-900045-4}{\emph{Eur. Phys. J. ST}
  {\bfseries 228} (2019) 261}.

\bibitem{FCC:2018vvp}
{\scshape FCC} collaboration, \emph{{FCC-hh: The Hadron Collider}: {Future
  Circular Collider Conceptual Design Report Volume 3}},
  \href{https://doi.org/10.1140/epjst/e2019-900087-0}{\emph{Eur. Phys. J. ST}
  {\bfseries 228} (2019) 755}.

\bibitem{Benedikt2020}
M.~Benedikt, A.~Blondel, P.~Janot, M.~Mangano and F.~Zimmermann, \emph{Future
  circular colliders succeeding the lhc},
  \href{https://doi.org/10.1038/s41567-020-0856-2}{\emph{Nature Physics}
  {\bfseries 16} (2020) 402}.

\bibitem{Diehl:1993br}
M.~Diehl and O.~Nachtmann, \emph{{Optimal observables for the measurement of
  three gauge boson couplings in e+ e- ---\ensuremath{>} W+ W-}},
  \href{https://doi.org/10.1007/BF01555899}{\emph{Z. Phys. C} {\bfseries 62}
  (1994) 397}.

\bibitem{DeBlas:2019qco}
J.~De~Blas, G.~Durieux, C.~Grojean, J.~Gu and A.~Paul, \emph{{On the future of
  Higgs, electroweak and diboson measurements at lepton colliders}},
  \href{https://doi.org/10.1007/JHEP12(2019)117}{\emph{JHEP} {\bfseries 12}
  (2019) 117} [\href{https://arxiv.org/abs/1907.04311}{{\ttfamily
  1907.04311}}].

\bibitem{Tweedie:2014yda}
B.~Tweedie, \emph{{Better Hadronic Top Quark Polarimetry}},
  \href{https://doi.org/10.1103/PhysRevD.90.094010}{\emph{Phys. Rev. D}
  {\bfseries 90} (2014) 094010}
  [\href{https://arxiv.org/abs/1401.3021}{{\ttfamily 1401.3021}}].

\bibitem{Frixione:2019lga}
S.~Frixione, \emph{{Initial conditions for electron and photon structure and
  fragmentation functions}},
  \href{https://doi.org/10.1007/JHEP11(2019)158}{\emph{JHEP} {\bfseries 11}
  (2019) 158} [\href{https://arxiv.org/abs/1909.03886}{{\ttfamily
  1909.03886}}].

\bibitem{Bertone:2019hks}
V.~Bertone, M.~Cacciari, S.~Frixione and G.~Stagnitto, \emph{{The partonic
  structure of the electron at the next-to-leading logarithmic accuracy in
  QED}}, \href{https://doi.org/10.1007/JHEP03(2020)135}{\emph{JHEP} {\bfseries
  03} (2020) 135} [\href{https://arxiv.org/abs/1911.12040}{{\ttfamily
  1911.12040}}].

\bibitem{Frixione:2012wtz}
S.~Frixione, \emph{{On factorisation schemes for the electron parton
  distribution functions in QED}},
  \href{https://doi.org/10.1007/JHEP07(2021)180}{\emph{JHEP} {\bfseries 07}
  (2021) 180} [\href{https://arxiv.org/abs/2105.06688}{{\ttfamily
  2105.06688}}].

\bibitem{Bertone:2022ktl}
V.~Bertone, M.~Cacciari, S.~Frixione, G.~Stagnitto, M.~Zaro and X.~Zhao,
  \emph{{Improving methods and predictions at high-energy $e^{+}e^{-}$
  colliders within collinear factorisation}},
  \href{https://doi.org/10.1007/JHEP10(2022)089}{\emph{JHEP} {\bfseries 10}
  (2022) 089} [\href{https://arxiv.org/abs/2207.03265}{{\ttfamily
  2207.03265}}].

\bibitem{Frixione:1993yw}
S.~Frixione, M.L.~Mangano, P.~Nason and G.~Ridolfi, \emph{{Improving the
  Weizsacker-Williams approximation in electron - proton collisions}},
  \href{https://doi.org/10.1016/0370-2693(93)90823-Z}{\emph{Phys. Lett. B}
  {\bfseries 319} (1993) 339}
  [\href{https://arxiv.org/abs/hep-ph/9310350}{{\ttfamily hep-ph/9310350}}].

\bibitem{Ruiz:2021tdt}
R.~Ruiz, A.~Costantini, F.~Maltoni and O.~Mattelaer, \emph{{The Effective
  Vector Boson Approximation in high-energy muon collisions}},
  \href{https://doi.org/10.1007/JHEP06(2022)114}{\emph{JHEP} {\bfseries 06}
  (2022) 114} [\href{https://arxiv.org/abs/2111.02442}{{\ttfamily
  2111.02442}}].

\bibitem{Shao:2022cly}
H.-S.~Shao and D.~d'Enterria, \emph{{gamma-UPC: automated generation of
  exclusive photon-photon processes in ultraperipheral proton and nuclear
  collisions with varying form factors}},
  \href{https://doi.org/10.1007/JHEP09(2022)248}{\emph{JHEP} {\bfseries 09}
  (2022) 248} [\href{https://arxiv.org/abs/2207.03012}{{\ttfamily
  2207.03012}}].

\bibitem{1405.0301}
J.~Alwall, R.~Frederix, S.~Frixione, V.~Hirschi, F.~Maltoni, O.~Mattelaer
  et~al., \emph{{The automated computation of tree-level and next-to-leading
  order differential cross sections, and their matching to parton shower
  simulations}}, \href{https://doi.org/10.1007/JHEP07(2014)079}{\emph{JHEP}
  {\bfseries 07} (2014) 079} [\href{https://arxiv.org/abs/1405.0301}{{\ttfamily
  1405.0301}}].

\bibitem{Frixione:2021zdp}
S.~Frixione, O.~Mattelaer, M.~Zaro and X.~Zhao, \emph{{Lepton collisions in
  MadGraph5\_aMC@NLO}},  \href{https://arxiv.org/abs/2108.10261}{{\ttfamily
  2108.10261}}.

\bibitem{Aguilar-Saavedra:2024hwd}
J.A.~Aguilar-Saavedra, \emph{{A closer look at post-decay $t \bar t$
  entanglement}},  \href{https://arxiv.org/abs/2401.10988}{{\ttfamily
  2401.10988}}.

\end{thebibliography}\endgroup
\end{document}